\documentclass[aps,prb,twocolumn,groupedaddress]{revtex4}

\usepackage{amsmath}
\usepackage{amsfonts}
\usepackage{siunitx}
\usepackage{graphicx}
\usepackage{bm}
\usepackage[caption=false]{subfig}
\graphicspath{ {./Images/} }

\newcommand{\ket}[1]{|#1\rangle}
\newcommand{\bra}[1]{\langle #1|}

\begin{document}

\title{Artificial Graphene in a Strong Magnetic Field: \\
       Bulk Current Distribution and Quantum Phase Transitions}

\author{Zeb E. Krix}
\email[]{z.krix@unsw.edu.au}
\author{Oleg P. Sushkov}
\affiliation{School of Physics, University of New South Wales, Sydney 2052, Australia}

\date{\today}

\begin{abstract}
We present calculations of the equilibrium current density and Chern numbers for a 2DEG in a periodic potential with infinite strip geometry and a perpendicular magnetic field. We consider a triangular lattice of anti-dots with large (a = 120 nm) lattice spacing. Such a system is known as artificial graphene (AG). To compute the current density we numerically diagonalise the AG Hamiltonian over a set of Landau level basis states, this takes into account coupling between different Landau levels. Our calculations show that, at magnetic fields typical for quantum Hall measurements, extended streams of current are present in the bulk of the sample. We investigate the scaling of these streams with potential strength. Knowledge of the AG energy levels allows us to compute the Chern number associated with each energy gap. We demonstrate that in tuning the height of the potential modulation the Chern number can undergo a transition between two different values.
\end{abstract}

\maketitle

\section{Introduction}

Artificial graphene (AG) is intended to simulate the electronic properties of graphene. The simulation must be controllable, including a possibility to switch on the spin orbit interaction. Such systems can be realised by imposing a super-modulation with triangular lattice-symmetry on a two dimensional gas of particles since this symmetry necessarily gives rise to Dirac cones. Artificial triangular arrays (superlattices) have been realised in semiconductor quantum wells with strong \cite{duEmergingManybodyEffects2018,wangObservationDiracBands2018,postTriangularNanoperforationBand2019} and weak\cite{geislerDetectionLandauBandCouplingInduced2004,iyeAharonovBohmtypeEffects2004} modulation, in twisted bilayer graphene\cite{forsytheBandStructureEngineering2018a,jessenLithographicBandStructure2019a}, and in optical lattices of cold atoms~\cite{jotzuExperimentalRealizationTopological2014a,uehlingerArtificialGrapheneTunable2013a}. For a review see Ref. \cite{poliniArtificialHoneycombLattices2013}

In the current work, we consider theoretically AG produced via electrostatic gating with repulsive lattice sites and typical lattice constants on the order of 100 nm. In principle the band structure of such a system should exhibit two sets of Dirac cones and a topological flat band \cite{tkachenkoEffectsCoulombScreening2015}. The latter work has established that arrays of anti-dots (when produced electrostatically) are less susceptible to disorder than arrays of dots. Due to the layered structure and due to the low energy scale spectroscopic methods such as ARPES are not applicable to these devices. Therefore, signatures of the superlattice potential are sought in transport measurements. More information can be obtained from transport properties when a perpendicular magnetic field is present and $R_{xx}$ and $R_{xy}$ are measured as functions of this field.

The problem of electron dynamics in superimposed magnetic field and periodic potential has been considered in numerous works. However, previously this problem was considered in the weak modulation limit, the amplitude of the potential modulation much smaller than the electron Fermi energy. For example in the experiment by Geisler et. al.\cite{geislerDetectionLandauBandCouplingInduced2004} the modulation amplitude was just  few per cent of the Fermi energy. Theoretically the weak modulation limit was addressed in Refs.\cite{hofstadterEnergyLevelsWave1976, claroMagneticSubbandStructure1979, thoulessQuantizedHallConductance1982,macdonaldQuantizedHallEffect1984a, gerhardtsNovelMagnetoresistanceOscillations1989}. However, for AG one needs the amplitude of the potential modulation to be several times larger than the Fermi energy, and this case has not been considered before. The strong modulation case can be addressed only numerically and in the present work we perform such a calculation considering an infinite stripe of a finite width. The finite width is necessary to address the edge states.

Of course in the weak modulation limit our results agree with previous works\cite{macdonaldQuantizedHallEffect1984a}. On the other hand, for strong modulation we observe two qualitatively new effects. (i) There are alternating 1D streams of current in the bulk. The amplitude of the stream current is proportional to the amplitude of the  modulation potential.  The streams can be detected by modern magnetometry techniques \cite{taylorHighsensitivityDiamondMagnetometer2008}. Of course there are also edge currents with amplitude independent of the modulation strength. (ii) We find that at a sufficiently large amplitude of the periodic modulation Chern numbers become different from those given by the weak modulation theory\cite{macdonaldQuantizedHallEffect1984a}. Hence, the sequence of quantum Hall plateaus of $R_{xy}$   depends on the modulation amplitude and we predict this dependence.

The problem we address is related to the integer quantum Hall effect without any supermodulation. The integer quantum Hall effect has been understood theoretically in terms of bulk\cite{stredaTheoryQuantisedHall1982} or edge\cite{halperinQuantizedHallConductance1982} currents. Experimental work reviewed in Ref.~\cite{weisj.MetrologyMicroscopicPicture2011} has investigated the spatial distribution of the Hall potential. This work indicates that under some conditions the quantum Hall  current is carried by both bulk and edge states. Of course in this case the bulk currents are due to disorder and impurities, not due to the periodically modulated potential.

The structure of the paper is the following: In section II we briefly discuss the details of this calculation. Section III discusses the fractal nature of the spectrum and presents calculations of the Chern number. In Section IV we discuss transitions in the Chern number which occur as potential strength is varied. And in Section V we discuss the presence of extended streams of current through the AG bulk and the way in which current density scales with potential strength.

\section{Matrix elements of the Hamiltonian and calculation techniques}
\begin{figure}
    \centering
    \includegraphics[scale=0.25]{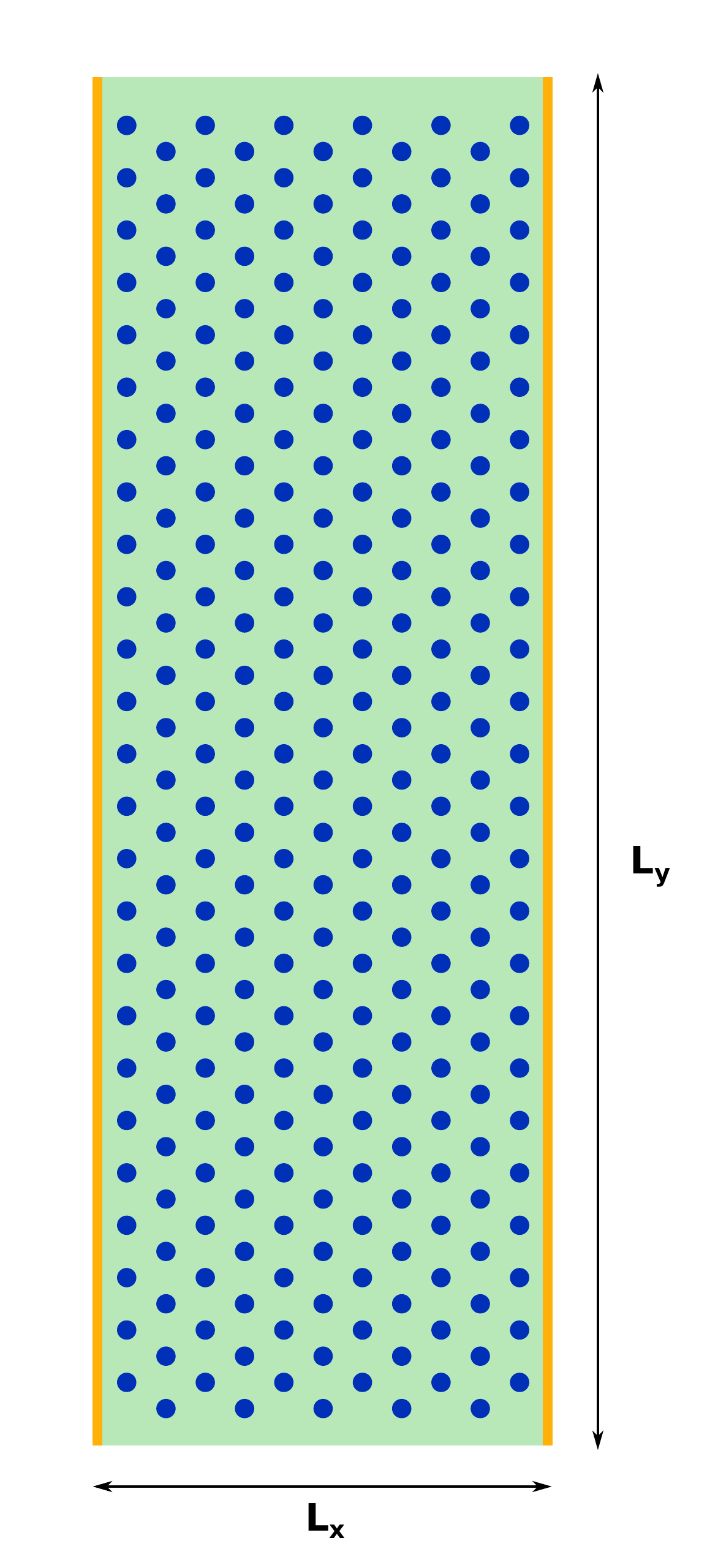}
    \caption{A long stripe of width $L_x=10a$ used in the calculations, $L_{x} \gg L_{x}$.}
    \label{fig:strip_geometry}
\end{figure}

The general approach used is the same as that in Ref \cite{macdonaldQuantizedHallEffect1984a} \cite{claroMagneticSubbandStructure1979}: A single particle Hamiltonian with triangular lattice potential is diagonalised over the basis of Landau level eigenvectors. We extend the calculation by computing the matrix elements of the Hamiltonian for arbitrary Landau levels and by allowing transitions between them. A further extension is the addition of a hard-walled confining potential $V(x)$ which defines the edges of the sample along the $x$-direction. This, together with periodic boundary conditions along the $y$-direction, imposes a strip geometry with very large aspect ratio (figure \ref{fig:strip_geometry}).

While the matrix elements of the lattice Hamiltonian can be computed exactly, the matrix elements of the confining potential must be computed via numerical integration. We can then numerically diagonalise this matrix to obtain energy levels (in the form of a dispersion relation) and eigenvectors. The remainder of this sections clarifies some of the details of this process:

Landau level eigenstates in the gauge $\bm{A} = (0, Bx, 0)$ are given by,

\begin{align}\label{equ:landaustates}
    \psi_{k,n}(x, y)
          &= A_{n} e^{iky} e^{-\xi^{2}/2} H_{n}(\xi) \\ \nonumber
    A_{n} &= \frac{1}{\sqrt{2^{n} n!}} \left( \frac{m \omega}{\pi} \right)^{1/4}
\end{align}

where $k$ is momentum along the length of the stripe, $x_{k} = k/eB$ is the center coordinate and $\xi = (x - x_{k})/l_{B}$ is position in units of the magnetic length ($l_{B} = 1/\sqrt{eB}$). The single-particle Hamiltonian is given by $H = \bm{p}^{2}/2m + U(\bm{r})$ with 

\begin{align}\label{equ:potential_form_1}
    U(\bm{r}) = 2W \sum_{i=1}^{3} \cos(g_{i} \cdot \bm{r})
\end{align}

for reciprocal lattice vectors $g_{i}$,

\begin{align}
    \bm{g}_{1} &= g_{0}(1/2,\sqrt{3}/2) \nonumber \\
    \bm{g}_{2} &= g_{0}(1,0) \\
    \bm{g}_{3} &= g_{0}(-1/2,\sqrt{3}/2) \nonumber \\
    \nonumber \\
   g_{0} = 2g/\sqrt{3},& \quad g = 2\pi /a \nonumber
\end{align}

We wish to compute the matrix elements,

\begin{align*}
    \bra{k_{i}, n} H \ket{k_{j}, m} =
    \delta_{nm} \delta_{ij} \omega_{c} ( n + 1/2 ) + \\
    \int \int \bar{\psi}_{k_{i},n}(\bm{r}-x_{k_{i}})
    \psi_{k_{j},m}(\bm{r}-x_{j}) (U(\bm{r}) + V(x)) \text{d} x \text{d} y
\end{align*}

With $\psi_{k,n}(\bm{r})$ given in Eq. \ref{equ:landaustates}. The integral over $U(\bm{r})$ can be evaluated analytically for arbitrary $n$, $m$. To do this we first write the potential (\ref{equ:potential_form_1}) in an alternate form,

\begin{align}\label{equ:potential_form_2}
    U(\bm{r}) = 2W\cos(g_{0} x) + 4W\cos(g y) \cos(g_{0} x /2)
\end{align}

The first of these terms is diagonal in $y$-momentum and the second is strictly off-diagonal in $y$-momentum, only mixing those states whose momenta are separated by $\pm g$. Performing the integration gives the following form for the matrix elements of $U$,

\begin{align}\label{equ:tri_ham}
    \bra{k_{i},n} &\hat{U} \ket{k_{j},m} = \\
    &2W\delta_{ij}X_{nm}(k_{i}) + \nonumber \\
    &2W\delta(k_{i} - k_{j} \pm g)
    [&&\cos(g_{0}x_{k}/2)F_{nm}^{(1,\pm)} \ + \quad \quad \quad \nonumber\\
&    &&\sin(g_{0}x_{k}/2)F_{nm}^{(2,\pm)} \ ] \quad \quad \quad \nonumber
\end{align}

where these two terms correspond to the two terms in Eq. \ref{equ:potential_form_2} respectively. We find the following equations for the matrices $X(k)$ and $F^{(i,\pm)}$:

\begin{widetext}
\begin{align}\label{equ:X_matrix}
    X_{nm}(k) =
        \begin{cases}
            (-1)^{(m-n)/2}\cos(g_{0}x_{k}) \\
            \quad \times
            \frac{\sqrt{2^{n}n!}}{\sqrt{2^{m}m!}}
            (g_{0}l_{B})^{m-n} e^{-(g_{0}l_{B})^{2}/4}
            L_{n}^{m-n} ( (g_{0}l_{B})^{2}/2 ),
                & m = n \mod 2 \\
            (-1)^{(m-n-1)/2}\sin(g_{0}x_{k}) \\
            \quad \times
            \frac{\sqrt{2^{n}n!}}{\sqrt{2^{m}m!}}
            (g_{0}l_{B})^{m-n} e^{-(g_{0}l_{B})^{2}/4}
            L_{n}^{m-n} ( (g_{0}l_{B})^{2}/2 ),
                & m \neq n \mod 2 
        \end{cases}
\end{align}

\begin{align}\label{equ:F_matrix}
    F_{nm}^{i,\pm} =
        A_{n}A_{m} e^{-\xi_{g}^{2}/2} \sum_{r,p=0}^{n,m} 
        \binom{n}{r} \binom{m}{p} 
        (\mp \xi_{g})^{r} (\pm \xi_{g})^{p}
        (A_{n-r}A_{m-p})^{-1} 
        \left. X_{n-r,m-p}(G_{i}) \right|_{g_{0} \mapsto g_{0}/2}
\end{align}
\end{widetext}

Where $L_{n}^{m-n}$ is a Laguerre polynomial and $\xi_{k} = x_{k}/l_{B}$ (the centre coordinate in units of magnetic length). In the second equation (Eq. \ref{equ:F_matrix}) we have used the notation $G_{i=1} = \pm g/2$ and $G_{i=2} = \pm g/2 + \pi e B / g_{0}$. We have written the matrix $F^{i,\pm}$ in terms of X (Eq. \ref{equ:X_matrix}) with $g_{0}$ replaced everywhere by $g_{0}/2$. In the limit of small coupling between Landau levels setting $n = m = 0$ reduces the Hamiltonian to that given by MacDonald\cite{macdonaldQuantizedHallEffect1984a}.

What remains is to compute the matrix elements of the confining potential $V(x)$ which defines the edges of the sample. We define $V(x)$ in terms of the Heaviside step function $\theta(x)$.

\begin{align}\label{equ:edge_potential_function}
    V(x) = E_{0}\theta(|x| - L_{x}/2)
\end{align}

The sample geometry imposed by this edge potential is sketched in Fig.~\ref{fig:strip_geometry}.

For the purposes of this calculation we set $E_{0} = 30$ meV and $L_{x}=10a$ or $30a$ where $a = 120$ nm. The height of the edge potential is well above the energy of any states considered in this work. Matrix elements (\ref{equ:edge_potential_matrix}) of the confining potential cannot be computed analytically. We therefore perform the integration numerically.

\begin{align}\label{equ:edge_potential_matrix}
    \bra{k,n} V \ket{k,m} =
        &E_{0}
        \int_{-\infty}^{x_{k} - L_{x}/2}
            \bar{\psi}_{k,n} (\bm{r}) \psi_{k,m}(\bm{r})
        \text{d}x \text{d}y \\
      + &E_{0}
        \int_{x_{k} + L_{x}/2}^{\infty}
            \bar{\psi}_{k,n}(\bm{r}) \psi_{k,m}(\bm{r})
        \text{d}x \text{d}y \nonumber
\end{align}

This completes our calculation of the Hamiltonian matrix.

With the Hamiltonian given we can numerically diagonalise this matrix to obtain all energy levels and eigenstates. The energy levels evolve with magnetic field in the well known Hofstadter butterfly pattern\cite{hofstadterEnergyLevelsWave1976}, with corrections due to having a triangular lattice (as opposed to a square one) and due to inter-Landau level coupling.

For the sake of convenience we measure magnetic field by the number of flux quanta per lattice unit cell. The flux quantum \footnote{Please note that this definition of the flux quantum differs from the superconducting flux quantum $h/2e$.} is $\phi_{0} = h/e$ and the flux per unit cell is $\phi = B a^{2} \frac{\sqrt{3}}{2}$. For the lattice spacing $a = 120$ nm that we consider

\begin{align}\label{equ:flux_to_tesla}
    \phi = \phi_0 \ \  \text{corresponds} \ \ \text{to} \ \ B=0.332 \ \text{T}.
\end{align}

The Hofstadter butterfly has a fractal structure. This band structure is simple when the flux is a rational number

\begin{align}\label{equ:rational_flux}
    \frac{\phi}{\phi_0} = \frac{p}{q}
\end{align}

In this case each Landau level is split to $p$ subbands \cite{claroMagneticSubbandStructure1979}.

In addition to the energy levels, numerical diagonalisation of the Hamiltonian gives a set of eigenstates $\psi_{\lambda,k}(\bm{r})$, where $\lambda$ enumerates energy bands of the combined magnetic field and superlattice system and $k$ is the $y$-momentum modulo $g$. As soon as the eigenstates are known we can compute the electric current.

\begin{align}\label{equ:current_denstity}
    j_{x,y} = \frac{ei}{2m} \sum_{\epsilon_{\lambda}(k) < \mu}
           ( &\psi_{\lambda,k}^{\dag}(\bm{r})
           [ \partial_{x,y} + i e A_{x,y} ] \psi_{\lambda,k}(\bm{r}) \\
           - &\psi_{\lambda,k}(\bm{r})
           [ \partial_{x,y} + i e A_{x,y} ] \psi_{\lambda,k}^{\dag}(\bm{r}) ) \nonumber
\end{align}

To find the total current density we sum contributions due to individual quantum states $\psi_{\lambda,k}(\bm{r})$ below the chemical potential (i.e $\psi_{\lambda,k}(\bm{r})$ such that $\epsilon_{\lambda}(k) < \mu$, where $\mu$ is the chemical potential). In practice we compute only the $y$-component of the current density (\ref{equ:current_denstity}) from the eigenstates and evaluate the $x$-component using the continuity equation.

\begin{align*}
    \partial_{x} j_{x} + \partial_{y} j_{y} & = 0 \\
    \implies j_{x}(x,y) & =
    \int_{-\infty}^{x} \partial_{y} j_{y}(x,y) \text{d} x \\
\end{align*}

The Hall conductivity can be calculated using the Streda equation\cite{stredaTheoryQuantisedHall1982}

\begin{align}\label{equ:streda}
    \sigma_{xy} = e \frac{\partial n_{s}}{\partial B}
\end{align}

where $n_{s}$ is the total electron density and the derivative is taken at constant chemical potential. Application of this equation requires counting the number of states below a fixed chemical potential and varying the magnetic field. It is valid only when the chemical potential is between subbands. In this case

\begin{align}\label{equ:def_chern_number}
  \sigma_{xy} = 2 \nu \frac{e^2}{h}
\end{align}

where $\nu$ is an integral Chern number. The factor 2 is for spin degeneracy which we assume throughout this work.

\section{Chern number calculation, Hall conductivity}
In this section we present results for the potential strength $W=0.4$ meV. The band width of this potential is

\begin{align}\label{equ:band_width}
    \Delta E = 9W = 3.6 \quad \text{meV}
\end{align}

For a magnetic field $B=1$ T the cyclotron frequency in GaAs is $\hbar \omega = 1.74$ meV and the capacity of a single Landau level (including spin degeneracy) is $n_L = 2\frac{B}{\phi_0} = 0.48 \times 10^{11} \text{cm}^{-2}$. Without a magnetic field this density would correspond to Fermi energy $\epsilon_{F} = 1.74$ meV. Both $\hbar \omega$ and $\epsilon_{F}$ are significantly smaller that the bandwidth (Eq. \ref{equ:band_width}). Therefore it seems that the potential (Eq. \ref{equ:band_width}) is strong. However, previous work \cite{tkachenkoEffectsCoulombScreening2015} has shown that for AG one needs a potential that is several times stronger. We are thus considering a weak to moderate potential.

\begin{figure}
    \centering
\subfloat
    [\label{fig:disp_phi=2.0}]
    {\includegraphics[width=0.22\textwidth]{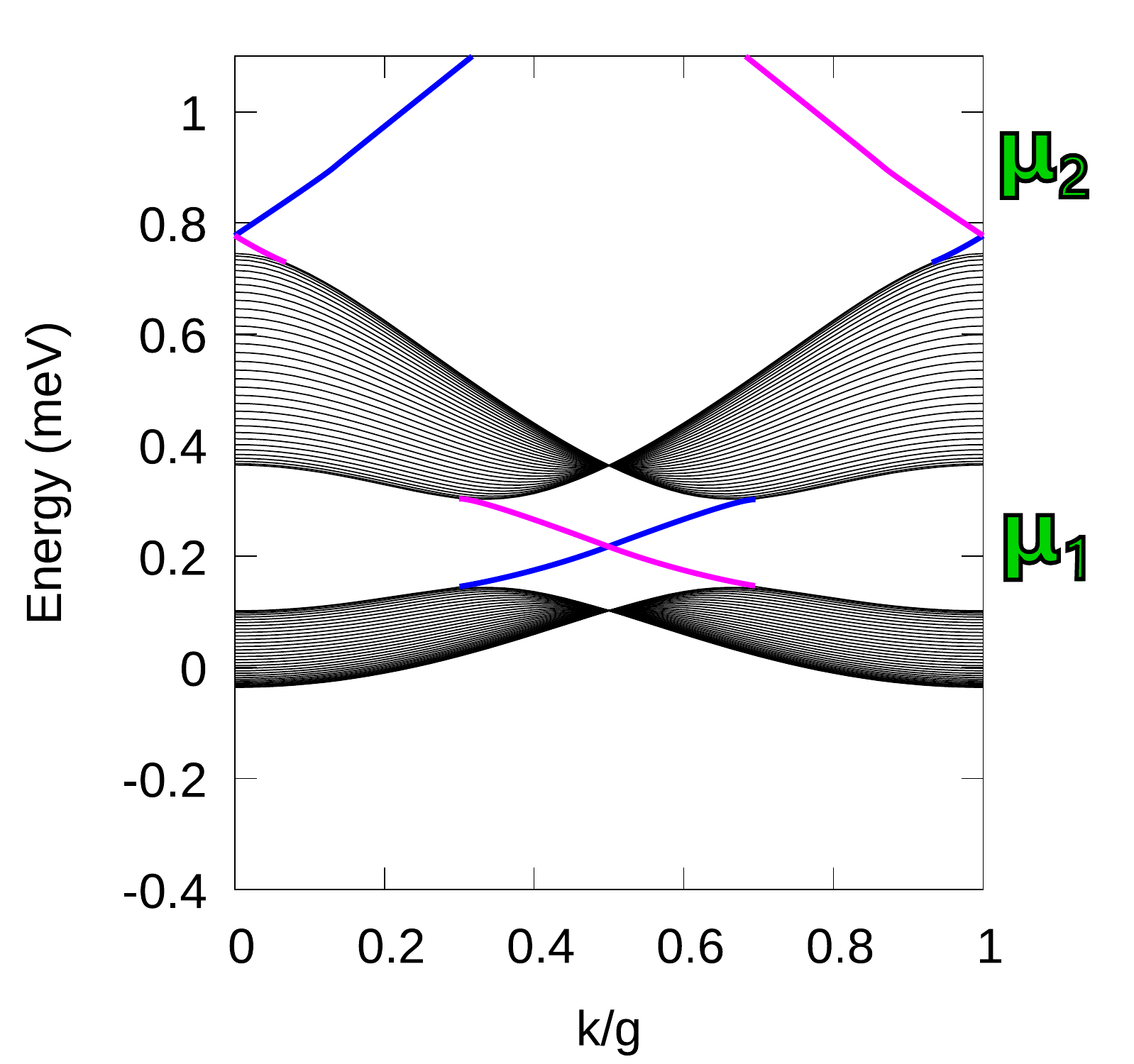}}
\subfloat
    [\label{fig:disp_phi=3.0}]
    {\includegraphics[width=0.22\textwidth]{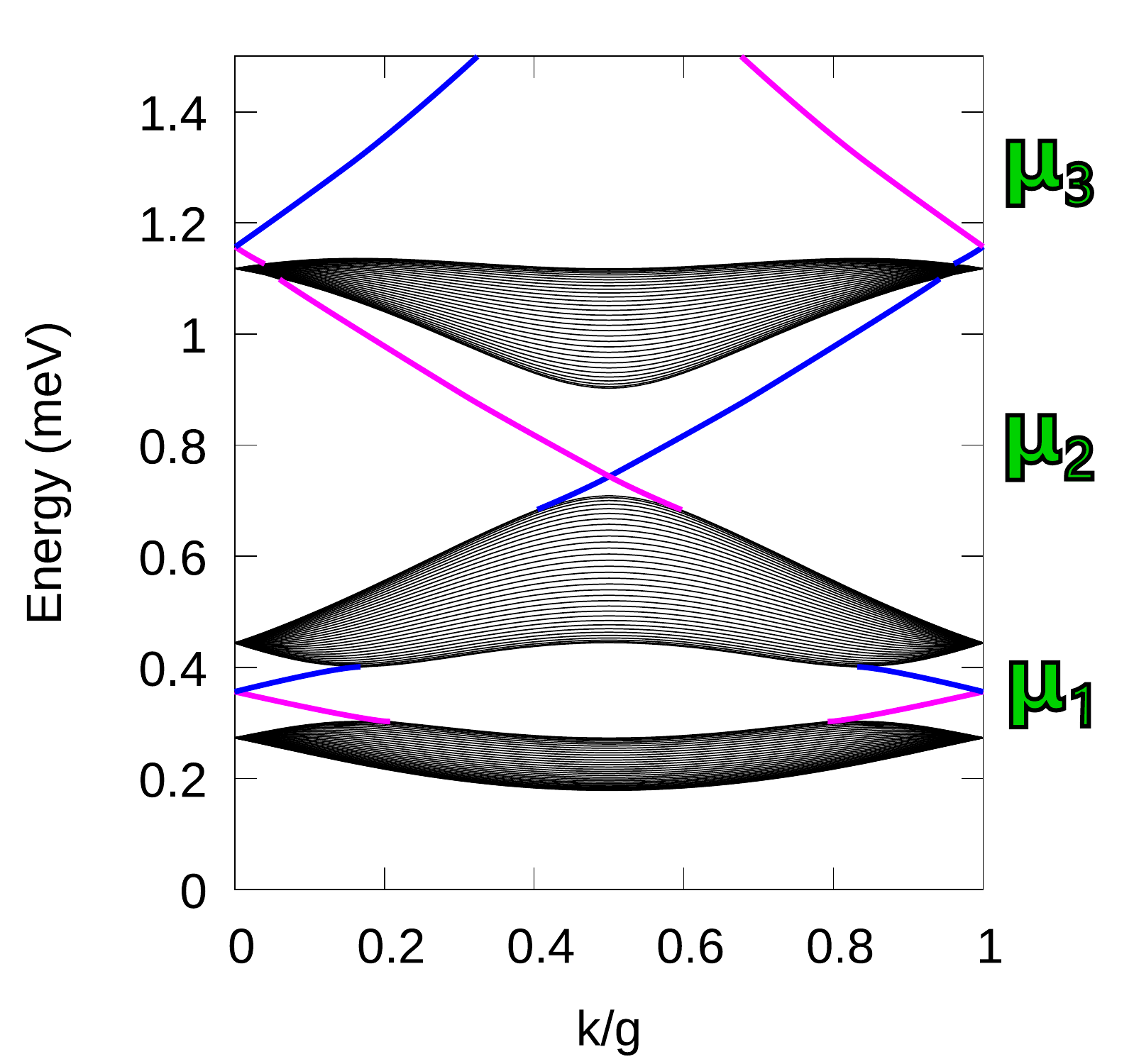}}

\subfloat
    [\label{fig:disp_phi=3.5}]
    {\includegraphics[width=0.22\textwidth]{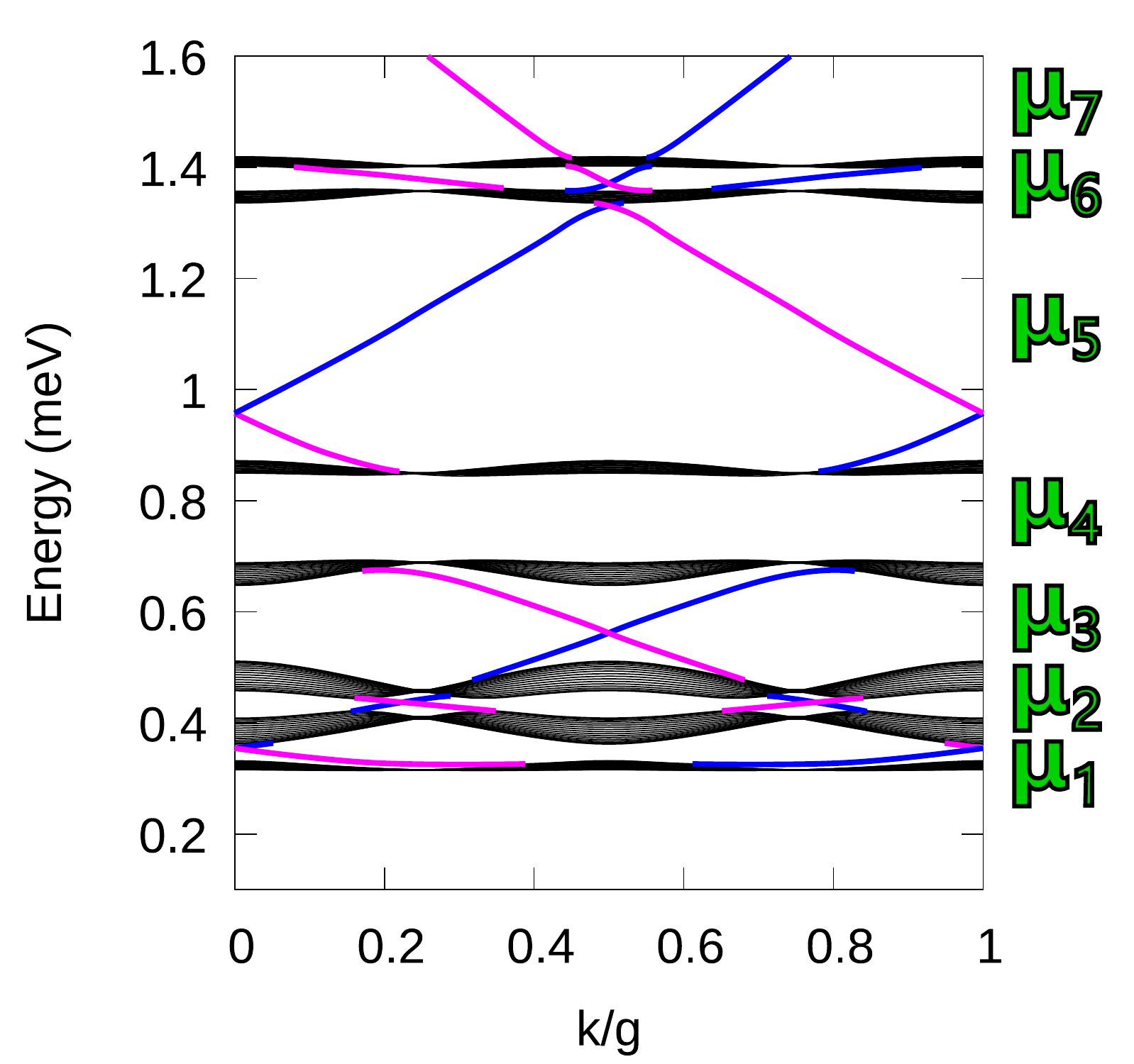}}
\subfloat
    [\label{fig:disp_phi=4.0}]
    {\includegraphics[width=0.22\textwidth]{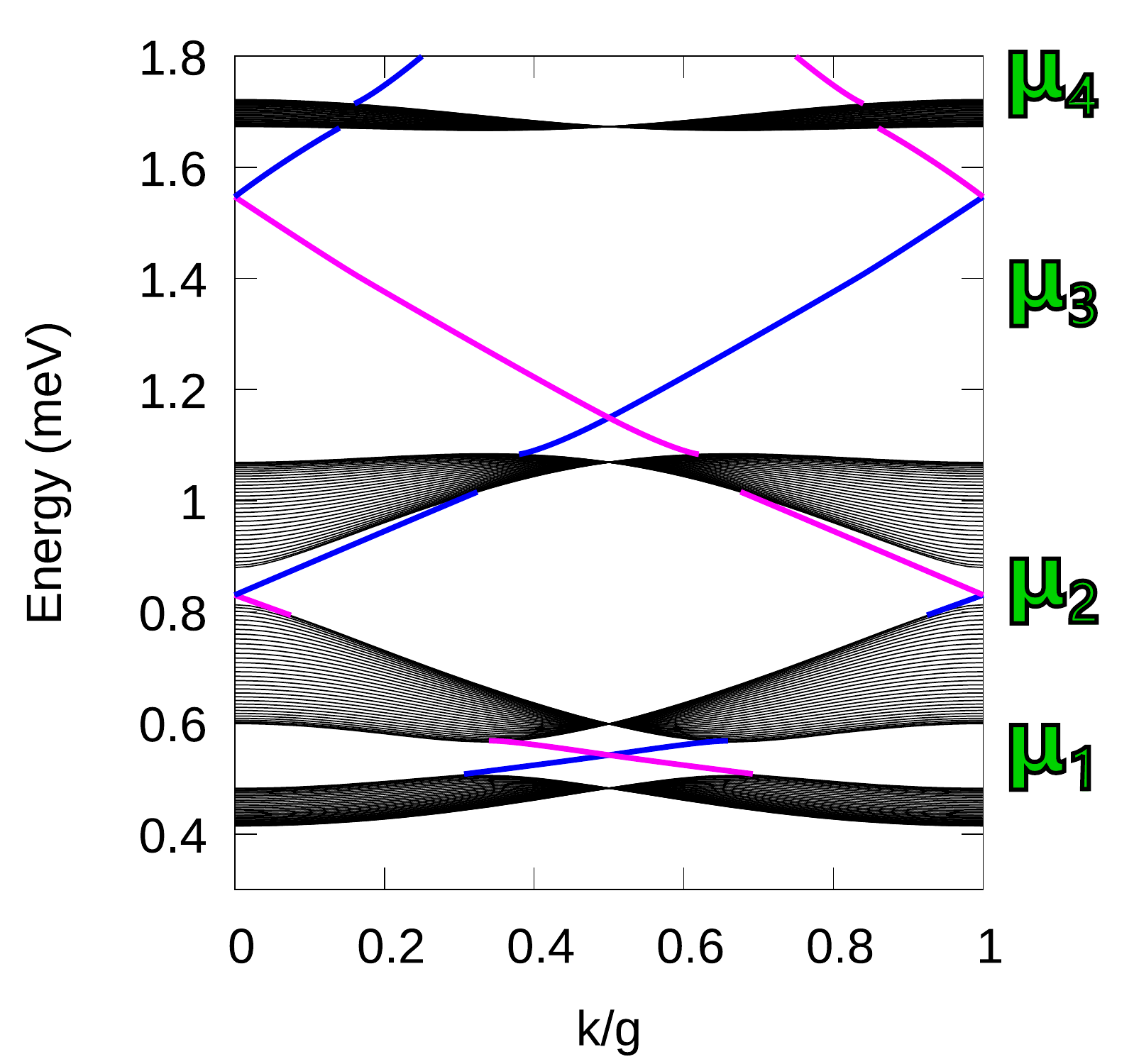}}
\caption{Calculated dispersions within the $n = 0$ Landau level for the following values of magnetic flux, (a) $\phi / \phi_0 = 2$ (2 subbands), (b) $\phi / \phi_0 = 3$ (3 subbands), (c) $\phi / \phi_0 = 7/2$ (7 subbands), (d) $\phi / \phi_0 = 4$ (4 subbands). The potential strength is $W = 0.4$ meV and the sample width is $L_{x} = 30a$ for superlattice period $a = 120$ nm. The edge state dispersions are shown in colour with the left edge coloured blue and the right edge coloured magenta.}
\label{fig:dispersion_all}
\end{figure}

In Fig. \ref{fig:dispersion_all} we present calculated dispersions for the split lowest Landau level at $\phi / \phi_0 = 2, \ 3, \ 7/2, \ 4$. These values of flux correspond to $B = 0.663, \ 0.995, \ 1.160, \ 1.326$ Tesla. From these figures it can be seen that the number of subbands is equal to the numerator of $\phi / \phi_{0}$.

For each case in figure \ref{fig:dispersion_all} the effective width of the $n = 0$ Landau level is about 1 meV. The lowest subband of the $n = 1$ Landau level (not shown in Fig. \ref{fig:dispersion_all}) is  about 1 meV above the topmost subband of the $n = 0$ Landau level. Thus the effective width of the Landau level is comparable to the separation between the levels, in this sense we have a weak to moderate strength potential.

The dispersions in Fig. \ref{fig:dispersion_all} consist of dense subbands of bulk states with edge states in ``gaps'' between the subbands. Since we know the wave functions we know to which edge, left or right, the edge state belongs (see Fig. \ref{fig:strip_geometry}). In Fig. \ref{fig:dispersion_all} the left edge dispersions are shown in blue and the right ones in magenta. Counting say the left edge states one can find the Chern number in Eq.(\ref{equ:def_chern_number}): in a given gap every left edge state with positive slope gives $\Delta\nu =1$, and every left edge state with negative slope gives $\Delta\nu =-1$. Hence the sequences of Chern numbers corresponding to the labelled values of chemical potential in Fig. \ref{fig:dispersion_all} are the following: panel (\ref{fig:disp_phi=2.0}) 1, 1.  Panel (\ref{fig:disp_phi=3.0}) 1, 1, 1. Panel (\ref{fig:disp_phi=3.5}) 1, 2, 1, 0, 1, 2, 1. Panel (\ref{fig:disp_phi=4.0}) 1, 1, 1, 1. As will discussed later in this section, these Chern numbers can be also calculated using the Streda equation (\ref{equ:streda}).

\begin{figure}[t]
\centering
\subfloat
    [\label{fig:dos_sig_phi=3.00}]
    {\includegraphics[width=0.48\textwidth]{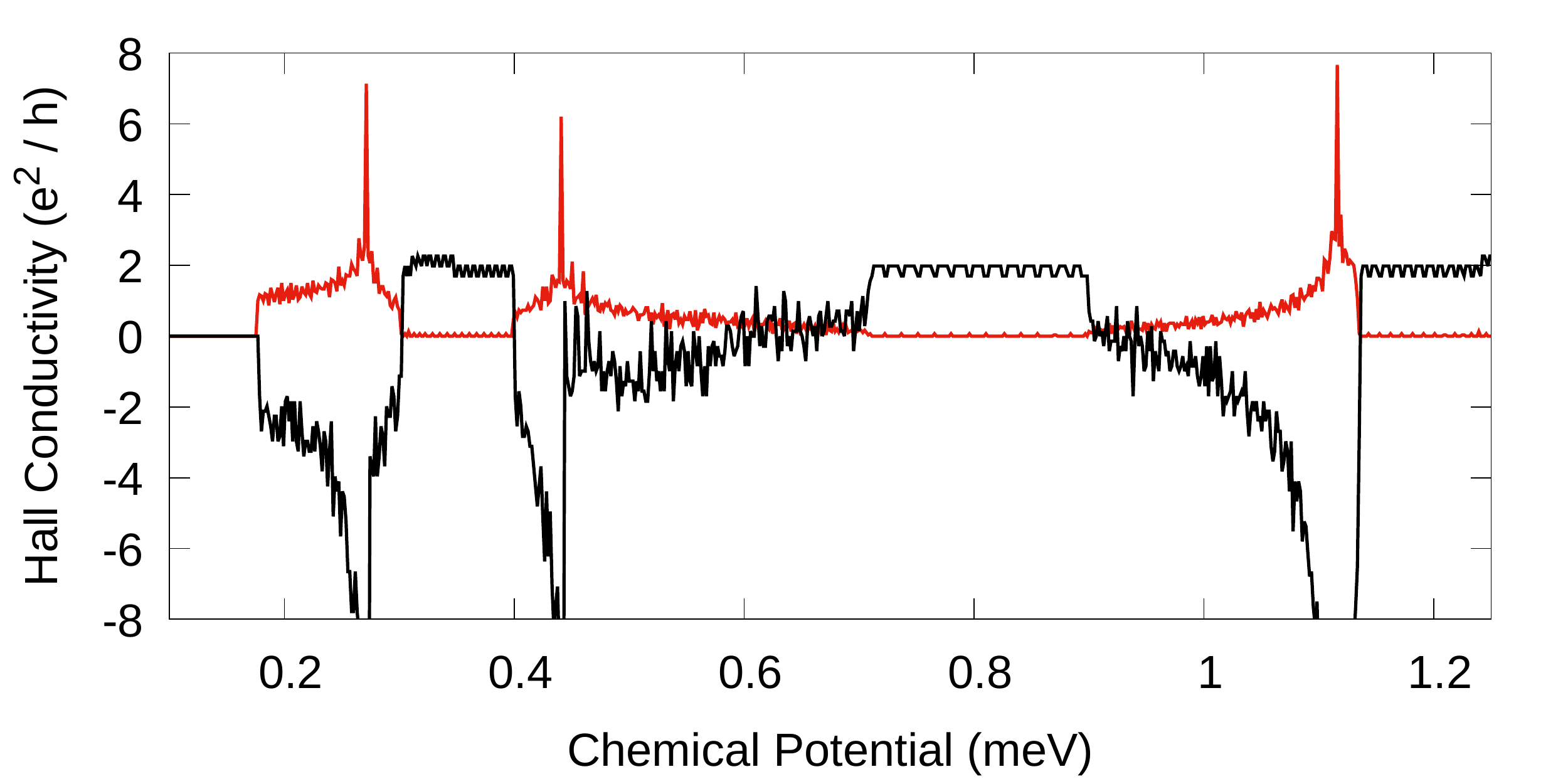}}

\subfloat
    [\label{fig:dos_sig_phi=2.98}]
    {\includegraphics[width=0.48\textwidth]{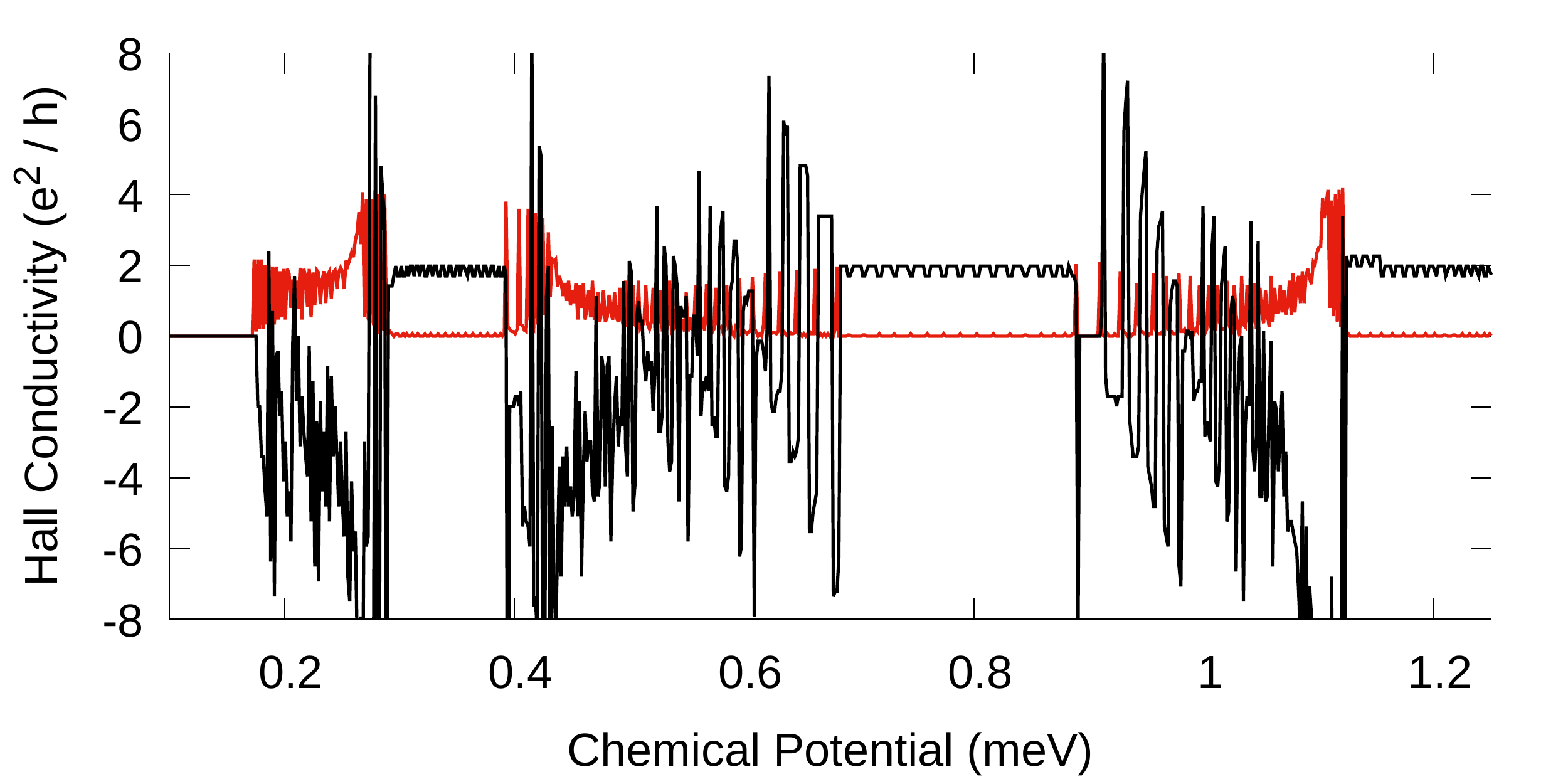}}

\subfloat
    [\label{fig:dos_sig_phi=2.96}]
    {\includegraphics[width=0.48\textwidth]{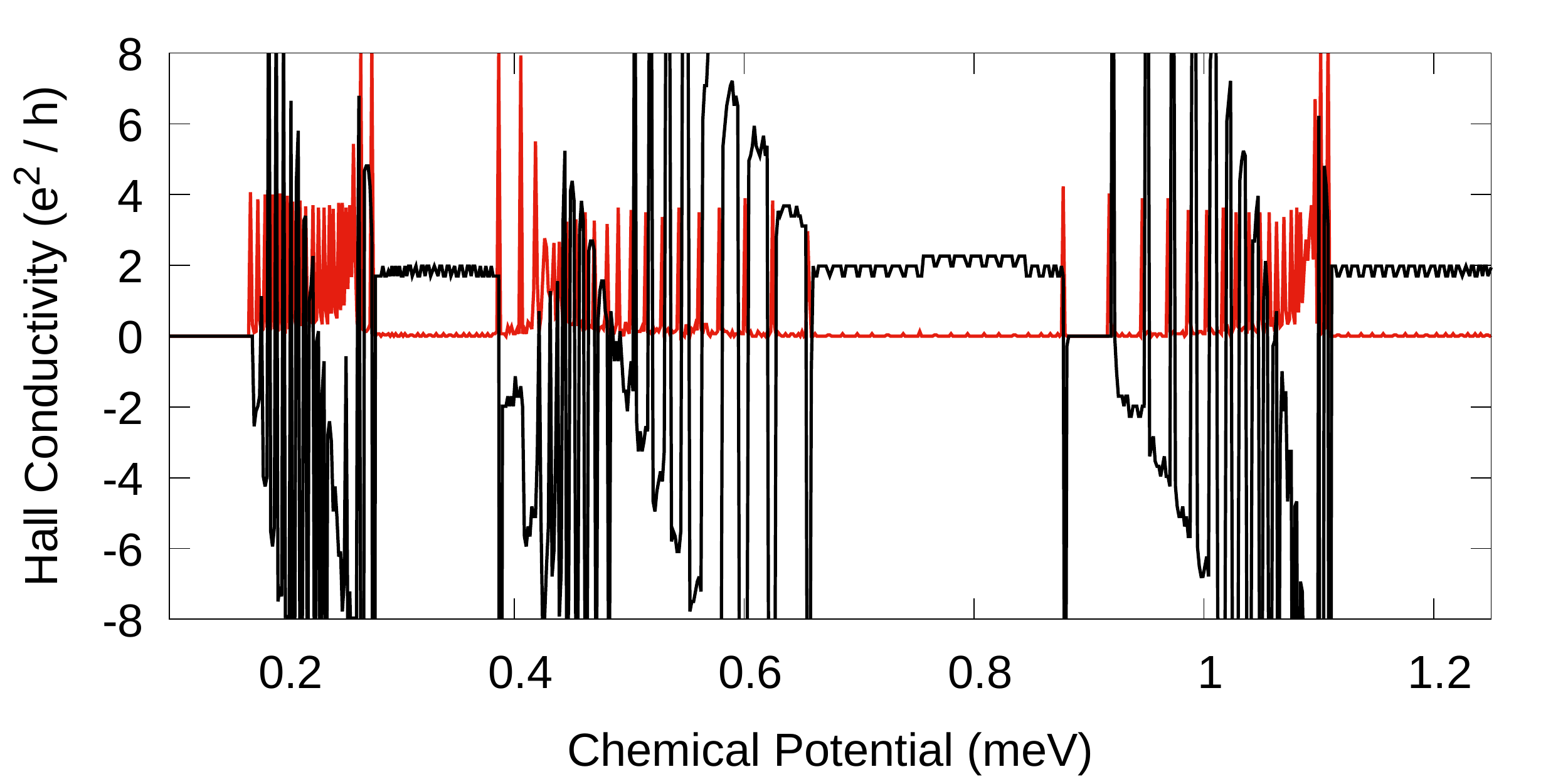}}
\caption{Hall conductivity (black) as a function of chemical potential for 3 different magnetic fields. Density of states (red) is overlaid in arbitrary units. As in figure \ref{fig:dispersion_all} we have $W = 0.4$ meV and $L_{x} = 30$ a. The values of magnetic flux used in each panel are: (a) $\phi / \phi_{0} = 3$, (b) $\phi / \phi_{0} = 2.98 = 149 / 50$ and (c) $\phi / \phi_{0} = 2.96 = 74 / 25$.}
\label{fig:dos_sig}
\end{figure}

Because of the fractal nature of the Hofstadter butterfly it is instructive to see how the band structure changes in the vicinity of  $\frac{\phi}{\phi_0} = \frac{p}{q}$. In Fig. \ref{fig:dos_sig} we plot the Hall conductivity (black) and the density of states (red) as a function of energy and consider fields close to 3 flux quanta. The density of states is given in arbitrary units. Panel (\ref{fig:dos_sig_phi=3.00}) corresponds to $\frac{\phi}{\phi_0} = 3$ and to figure \ref{fig:disp_phi=3.0}, panel (\ref{fig:dos_sig_phi=2.98}) to $\frac{\phi}{\phi_0} = 2.98 = \frac{149}{50}$ and panel (\ref{fig:dos_sig_phi=2.96}) to $\frac{\phi}{\phi_0} = 2.96 = \frac{74}{25}$. Figure \ref{fig:dos_sig_phi=3.00} shows clearly that there are 3 subbands. In the gaps between bulk bands when the density of states goes to zero, the Hall conductivity is $\sigma_{xy} = 2 e^{2} / h$. Inside the bands it takes some fluctuating negative values. The calculated values of $\sigma_{xy}$ inside the bands does not have any physical meaning since here Eq. (\ref{equ:streda}) is not valid.

Small changes in the magnetic field give a dramatic change in the numerator of $\phi / \phi_{0}$. Figure \ref{fig:dos_sig} illustrates how this is possible. A change in field of around $1 \%$ from figure \ref{fig:dos_sig_phi=3.00} to figure \ref{fig:dos_sig_phi=2.96} changes the number of subbands from 3 to 74. On the other hand the broad gaps in figures \ref{fig:dos_sig} are almost identical. The new narrow gaps in panel (\ref{fig:dos_sig_phi=2.96}) ``arise'' from the continuous spectrum in panel (\ref{fig:dos_sig_phi=3.00}). In this sense there is some ``continuity'' in the fractal butterfly.

\begin{figure}[t]
    \caption{Colour map of $\sigma_{xy}$ computed numerically using equation 
\ref{equ:streda} plotted as a function of chemical potential and flux per unit cell. As in figure \ref{fig:disp_phi=3.0} the potential strength is $W = 0.4$ meV  and only the lowest Landau level is shown.}
    \centering
    \includegraphics[width=0.47\textwidth]{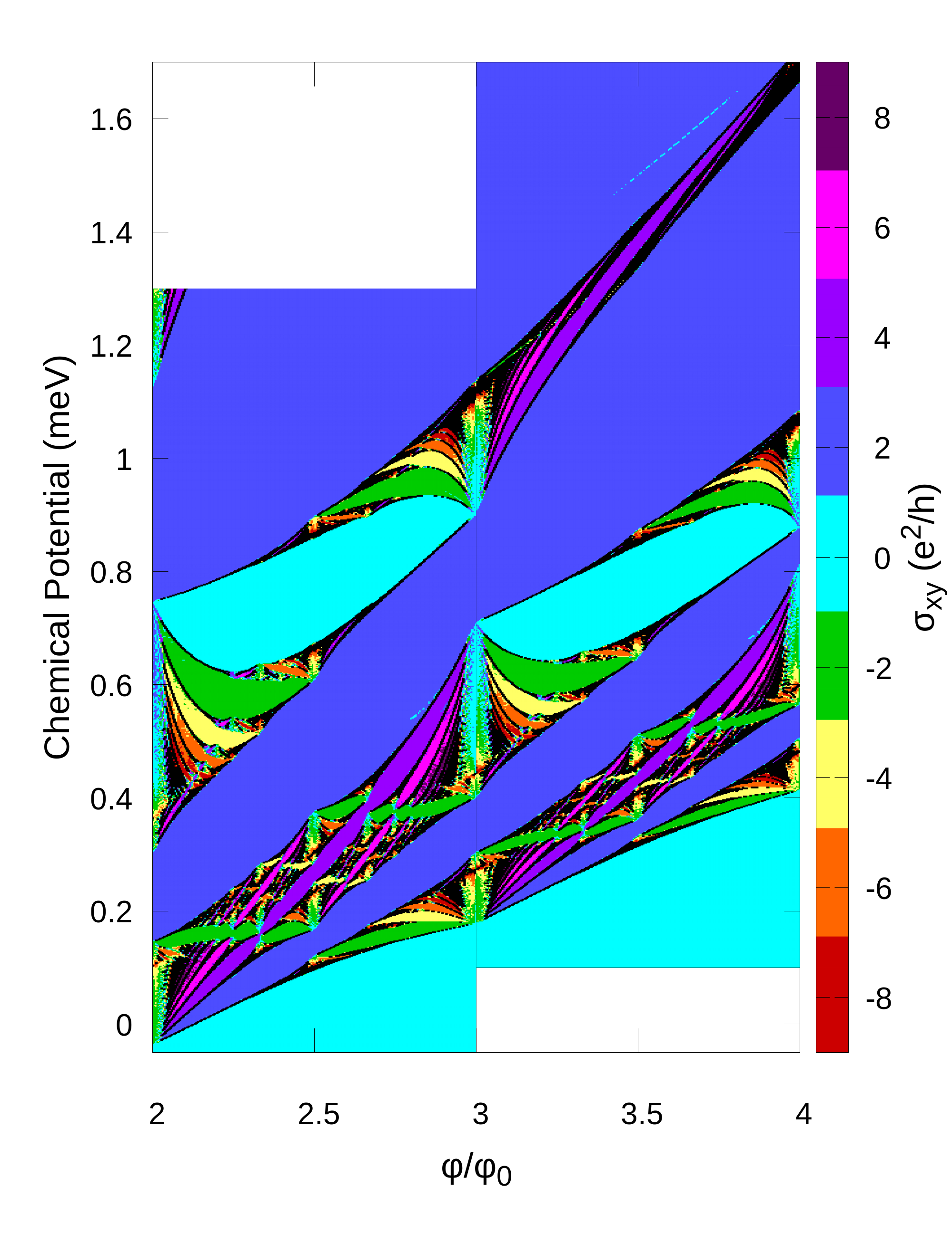}
    \label{fig:butterfly}
\end{figure}

We now present (in Fig. \ref{fig:butterfly}) the map of $\sigma_{xy}$ over the flux-energy plane. Strictly speaking a fractal function cannot be mapped in this way because all features in the spectrum that are smaller than some energy scale will not be represented. However, due to disorder and a finite sample size,  the Hofstadter butterfly of any real sample will not be a true fractal.

Once a finite sample width has been defined there is a limit on the maximum number of subbands that can be resolved (e.g. close to an integer flux). This limit is equal to the number of energy bands ($\Lambda$) in a single Landau level. If $N$ is the total number of states within a Landau level then

\begin{align}\label{equ:num_energy_bands}
    N = n_{s} L_{x} L_{y} &= 2 \Lambda g L_{y} / 2 \pi \nonumber \\
                          &= 2 B L_{x} L_{y} / \phi_{0} \nonumber \\
         \implies \Lambda &= B L_{x} a / \phi_{0}
\end{align}

Where we have used the fact that a single energy band contains a number of states equal to $2 \int d\text{k}L_{y}/2 \pi = 2 g L_{y} / 2 \pi = 2 L_{y} / a$. In the case of figure \ref{fig:dos_sig} with $L_{x} = 30a$, equation \ref{equ:num_energy_bands} gives $\Lambda = 104$. The implication is that all subbands should be visible in \ref{fig:dos_sig_phi=3.00} and \ref{fig:dos_sig_phi=2.96} while in figure \ref{fig:dos_sig_phi=2.98} it is impossible to resolve the expected 149 subbands. Disorder will also introduce a broadening of the density of states which obscures details of the fractal structure.

For the sake of visual clarity and to eliminate points in the $(\mu, B)$ plane for which Eq. \ref{equ:streda} is not valid we imitate this loss of fine structure in figure \ref{fig:butterfly}. To do this we note that in figure \ref{fig:dos_sig} when $\mu$ is inside a bulk band $\sigma_{xy}$ takes large negative values. We thus colour all points in the map \ref{fig:butterfly} black if the Hall conductivity falls outside of the region $\sigma_{xy} \in [-8, 8]$. This procedure eliminates a large number of the unphysical points, however it does not register the bulk bands at simple rational fluxes (e.g. $\phi = 2, \ 3, \ 4$).

Thus figure \ref{fig:butterfly} shows in colour the Hall conductivity of only the largest energy gaps and shows in black the bulk subbands of the Hofstadter butterfly. We make predictions for the sequence of $\sigma_{xy}$ values at arbitrary fields in the range $\phi / \phi_{0} \in [2, 4]$. For example, we compute the following sequences: at $\phi / \phi_{0} = 5 / 2$ $\sigma_{xy} = $ 2, 4, 2, 0, 2, at $\phi / \phi_{0} = 8 / 3 = 2.\dot{6}$ $\sigma_{xy} = $ -2, 2, 6, 4, 2, 0, -2, and at $\phi / \phi_{0} = 10 / 3 = 3.\dot{3}$ $\sigma_{xy} = $ 2, -2, 6, 2, -2, 0, 2, 4, 6, 2. Previous work by MacDonald \cite{macdonaldQuantizedHallEffect1984a} has computed these sequences in the small $W$ limit and in the highly anisotropic limit (here, anisotropic means that the second term in the potential (\ref{equ:potential_form_2}) is much smaller than the first). At small $W$ there are discrepancies with our results due to the change from an anisotropic to isotropic lattice. For example, the second gap at $\phi / \phi_{0} = 7 / 3$ from -2 to 4 changes during this transition. At the values of $W$ we consider (where first order perturbation theory is no longer applicable) there are additional discrepancies which will be discussed in the next section. All of the values for $\sigma_{xy} = 2 \nu e^{2} / h$ that we observe are solutions to the Diophantine equation introduced by Thouless et. al.\cite{thoulessQuantizedHallConductance1982}. If $i$ is the energy gap index ($i = 1, \ 2, \ \cdots, p$), $\phi / \phi_{0} = p / q$ and $w$ is an arbitrary integer then:

\begin{align}\label{equ:diophantine}
i = p \nu + q w
\end{align}

for every gap we observe. For example, in gap 1 at $\phi / \phi_{0} = 5 / 2$ we find $\nu = 1$ (Fig. \ref{fig:butterfly}). Equation \ref{equ:diophantine} is satisfied with $\nu = 1$ and $w = -2$: $1 = 5 \times 1 + 2 \times (-2)$. In this case equation \ref{equ:diophantine} forbids solutions $\nu = 0, \pm 2, \pm 4, \cdots$ since increasing $\nu = 1$ by unity would give an odd change to the first term which cannot be cancelled by the second term (which is even).

It is relevant to the results below that different integer solutions $\nu$ to the above equation are separated by $q$: Given some solution ($\nu$, $w$) to Eq. \ref{equ:diophantine} we can write $\nu = (i - q w) / p$. A second integer solution ($\nu'$, $w'$) can then be obtained by taking $w' = w - p$ so that $\nu' = \nu + q$.

As an illustration of Eq. \ref{equ:streda} consider the following example. When the chemical potential $\mu$ is in an energy gap between two Landau levels ($\mu_{3}$ in figure \ref{fig:disp_phi=3.0}) the number of states below $\mu$ is just an integer multiple, $\nu$, of the Landau degeneracy factor $n_{L} = 2 B / \phi_{0}$. Thus, $n_{s} = 2 e \nu B / h$. Computing $\sigma_{xy}$ via equation \ref{equ:streda} gives $\sigma_{xy} = 2 \nu e^{2}/h$, an integer multiple of $e^{2}/h$ as expected. For energy gaps \textit{inside} a Landau level (e.g. $\mu_{1}$, \ $\mu_{2}$ in figure \ref{fig:disp_phi=3.0}) the calculation can similarly be done by hand.

Figure \ref{fig:butterfly} shows the function $\sigma_{xy}(\mu,B)$ computed via equation \ref{equ:streda} with $n_{s}$ obtained from the set of energy levels that come out of the numerical diagonalisation procedure. Since the density of states is dominated by bulk states this can be considered a bulk calculation. This figure (in addition to showing $\sigma_{xy}$) illustrates the evolution of the spectrum as magnetic field is varied. For example it can be seen that at $\phi / \phi_{0} = 2$, 3, 4 there are 2, 3, and 4 subbands respectively. As an example of how equation \ref{equ:streda} works \textit{within} a Landau level consider the gap in figure \ref{fig:butterfly} which continuously stretches from gap 1 at $\phi = 2$ ($\mu_{1}$ in Fig. \ref{fig:disp_phi=2.0}) to gap 3 at $\phi = 4$ ($\mu_{3}$ in Fig. \ref{fig:disp_phi=4.0}). Figure \ref{fig:butterfly} shows that these two gaps are connected by a continuous region in the $(\mu,B)$ plane that contains only edge states. Looking only at the structure of the energy levels the former gap has $n_{1}$ states below and the latter has $n_{2}$ states below where

\begin{align}
    n_{1} &= \frac{1}{3} n_{L} &&= 2 \frac{1}{3} \frac{B}{\phi_{0}}
          &&= 2 \frac{1}{3} n_{0} \frac{\phi}{\phi_{0}} &&= 2 n_{0} \\
    n_{2} &= \frac{3}{4} n_{L} &&= 2 \frac{3}{4} \frac{B}{\phi_{0}}
          &&= 2 \frac{3}{4} n_{0} 4 &&= 6 n_{0}
\end{align}

We have used the proportionality constant $n_{0} = (2/\sqrt{3})a^{-2} = 8.02 \times 10^{9}$ $\text{cm}^{-2}$ between $B$ and $\phi$. Crucial to the above calculation is the fact that when $\phi = (p/q)\phi_{0}$ the number of states within a Landau level is evenly distributed between the $p$ subbands. This fact holds within our numerical calculation and can also be demonstrated analytically (see e.g. Ref.\cite[Chapter 12.9]{girvin_yang_2019} for an argument using the square lattice which applies equally well to triangular lattices). Thus the change in density between these two points is $\Delta n = 4n_{0}$ while the change in magnetic field is $\Delta B = n_{0} \Delta \phi = 2n_{0} \phi_{0}$. This calculation gives a Hall conductivity of

\begin{align*}
    \sigma_{xy} = e \frac{\Delta n_{s}}{\Delta B} =
                  e \frac{4n_{0}}{2n_{0}\phi_{0}} =
                  2 \frac{e^{2}}{h}
\end{align*}

which agrees with the numerical result in figure \ref{fig:butterfly}. It also agrees with the results discussed above for Fig. \ref{fig:disp_phi=3.0} and Fig. \ref{fig:disp_phi=4.0}, obtained by counting edge modes.

\section{Quantum Phase Transitions Driven by Potential Strength}

\begin{figure}[t]
\centering
\subfloat
    [\label{fig:transtion_W=0.5}]
    {\includegraphics[width=0.3\textwidth]{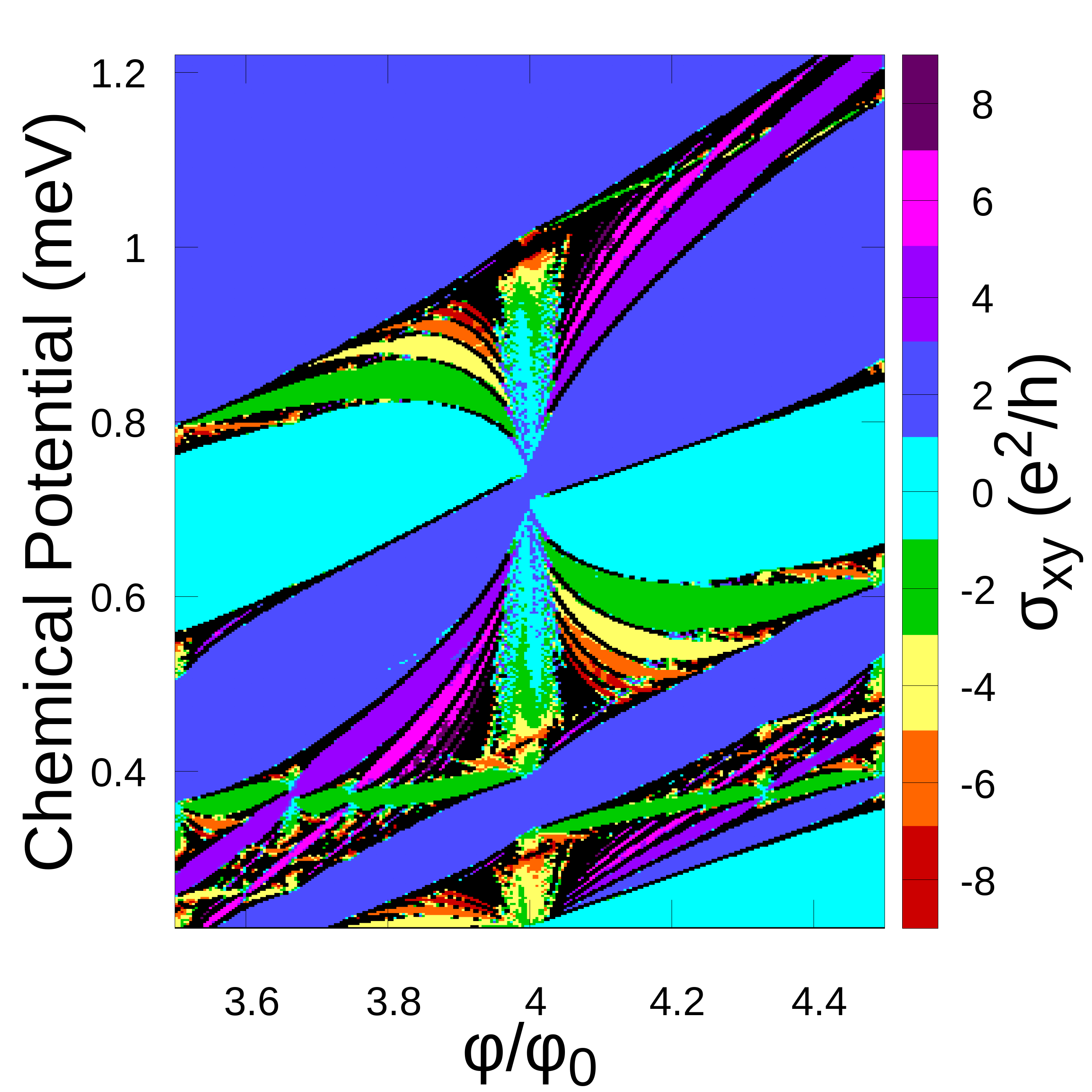}}

\subfloat
    [\label{fig:transtion_W=0.6}]
    {\includegraphics[width=0.3\textwidth]{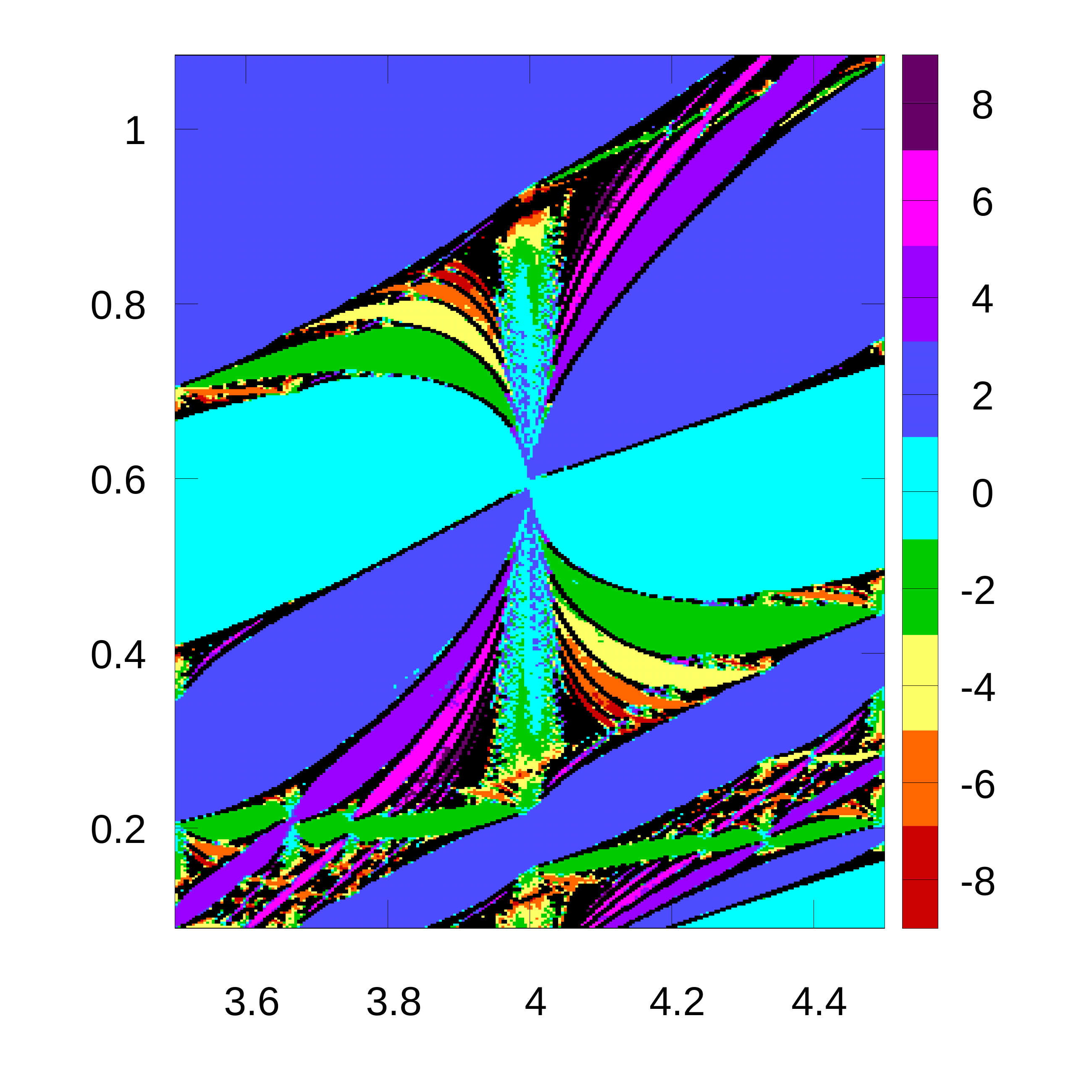}}

\subfloat
    [\label{fig:transtion_W=0.7}]
    {\includegraphics[width=0.3\textwidth]{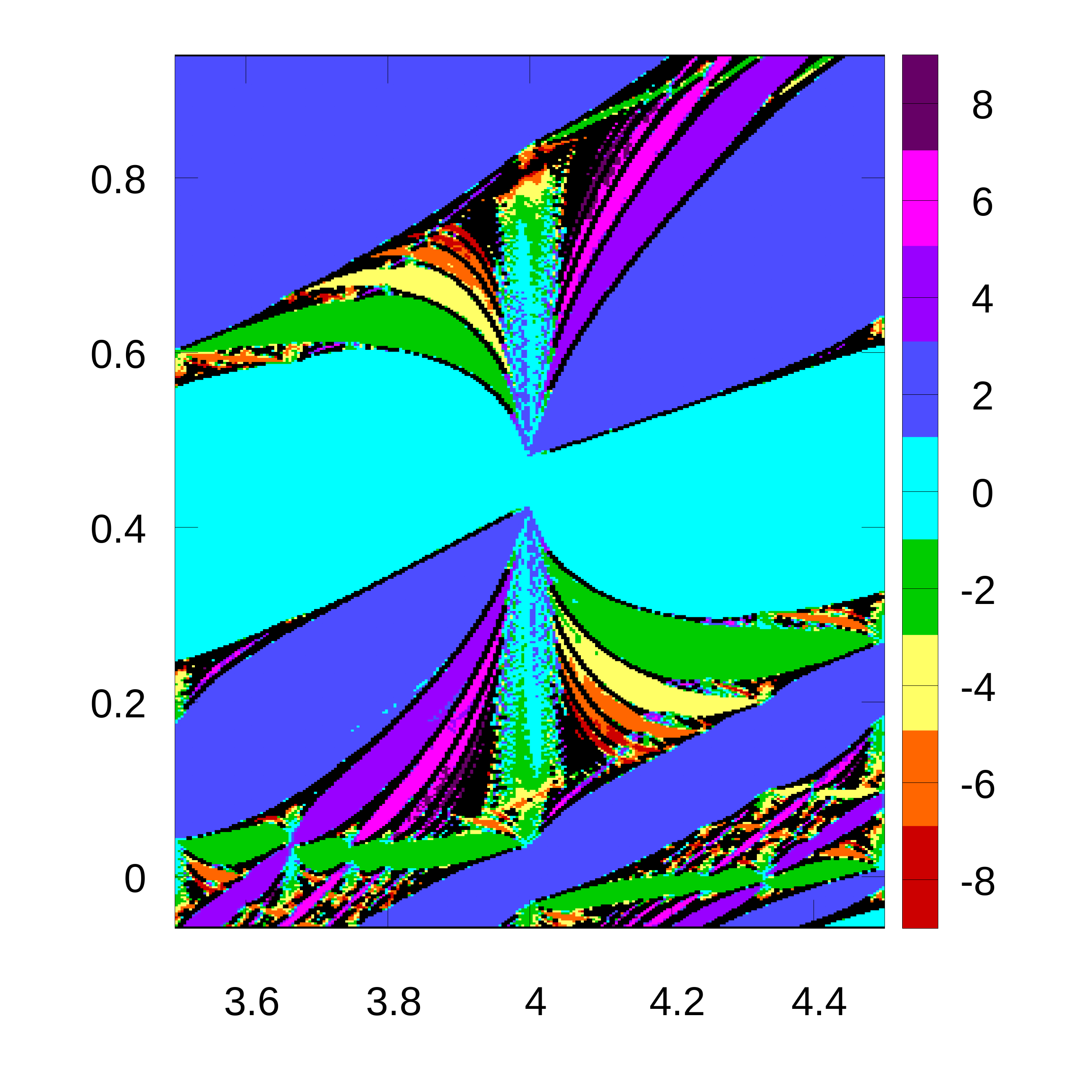}}
\caption{Maps of $\sigma_{xy}$ over the $(\mu, B)$ plane in the region around gap 2 at $\phi / \phi_{0} = 4$. Each plot shows this region at a different potential strength $W$. Panel (a) corresponds to $W = 0.5$ meV, panel (b) to $W = 0.6$ meV and panel (c) to $W = 0.7$ meV.}
\label{fig:transition_W}
\end{figure}

In calculations of Hall conductivity (figure \ref{fig:butterfly}) at different potential strengths we have observed energy gaps whose Chern numbers change as $W$ is increased.  Previous numerical work \cite{springsguthHallConductanceBloch1997} has found changes in the Chern number for a \textit{square} lattice in a magnetic field as coupling between Landau levels is introduced. Chern numbers change when a given energy gap closes and then reopens as a function of some parameter\cite{kaneQuantumSpinHall2005}; in this case $W$. An example of such a transition is given in figure \ref{fig:transition_W} which focuses on gap 2 at $\phi / \phi_{0} = 4$. In this figure we present an identical calculation to that in figure \ref{fig:butterfly} except over the field range $\phi / \phi_{0} = 3.5$ to 4.5 and for a larger potential $W = 0.5, \ 0.6, \ 0.7$ meV. The first panel \ref{fig:transtion_W=0.5} corresponds to $W = 0.5$ meV and there is one large continuous gap with Chern number 1 (coloured dark blue). There are also two large disconnected gaps with Chern number 0 (coloured light blue). Increasing the potential to $W = 0.6$ meV in figure \ref{fig:transtion_W=0.6} closes the energy gap and gives 4 disconnected regions. Increasing $W$ further to $W = 0.7$ meV reopens the energy gap in such a way that the Chern number within the gap is now 0. The two light blue regions from figure \ref{fig:transtion_W=0.5} have become connected and the single dark blue region has become disconnected.

\begin{figure}[t]
    \caption{Schematic of the evolution in band structure on the $(\mu,B)$ plane which leads to a change in Chern number. The black bands represent bulk states and the white and blue regions represent connected regions of edge states. As $W$ is increased the central energy gap (at $\phi / \phi_{0} = q / p$) closes and then reopens with a different Chern number. Chern numbers in each energy gap are labelled $\sigma_{i}$. We note that if the central gap occurs at a flux $\phi / \phi_{0} = p / q$ then the difference in Chern number between the two gaps shown is an integer multiple of $q$.}
    \centering
    \includegraphics[width=0.3\textwidth]{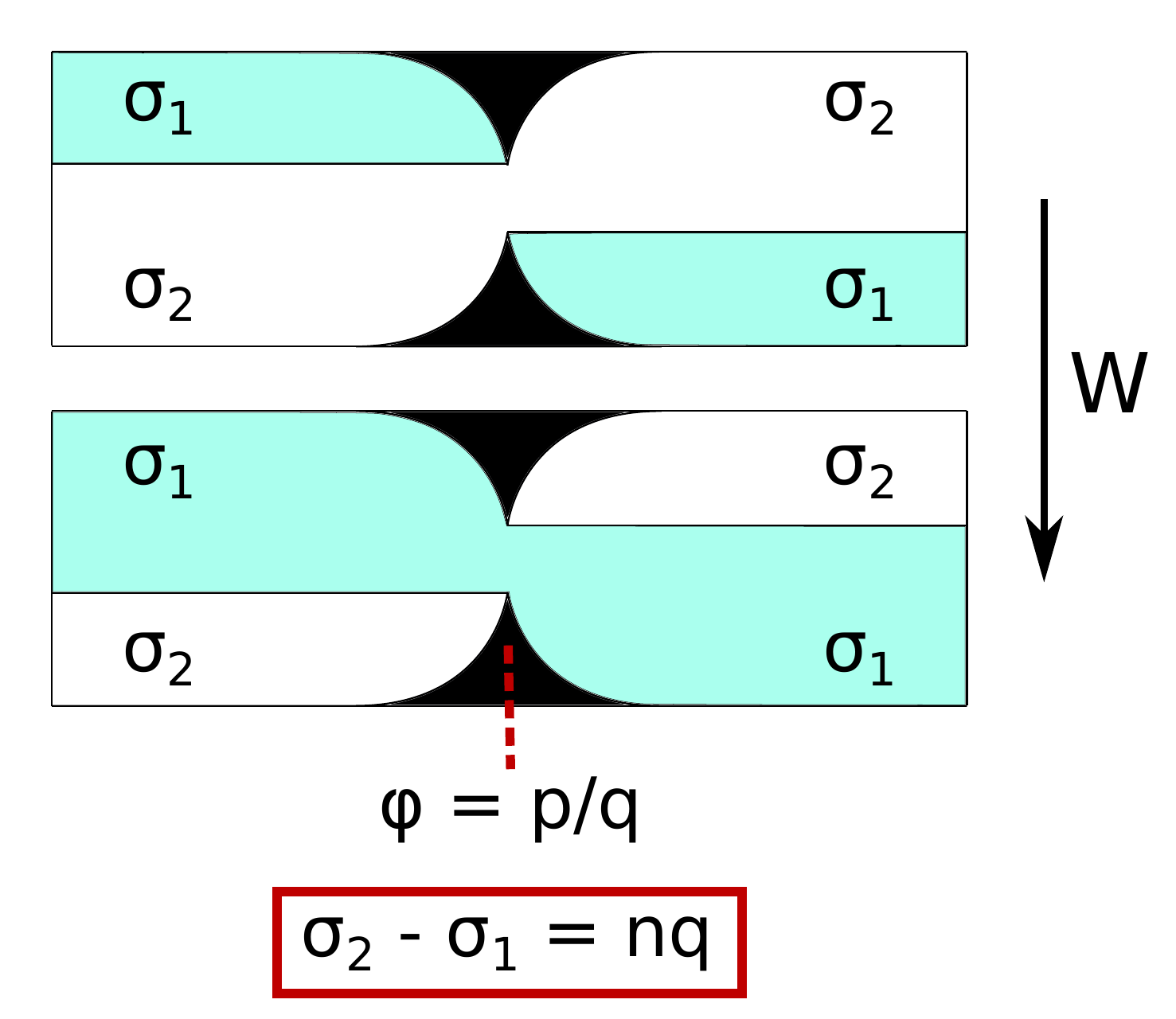}
    \label{fig:butterfly_trans_schem}
\end{figure}

Figures \ref{fig:transtion_W=0.5} and \ref{fig:transtion_W=0.7} are represented schematically in figure \ref{fig:butterfly_trans_schem}. In this case $\sigma_{1} = 0$ and $\sigma_{2} = 1$. The top panel in figure \ref{fig:butterfly_trans_schem} shows a region of the $(\mu, B)$ plane for a value of $W$ before the transition (figure \ref{fig:transtion_W=0.5}) while the bottom panel shows that same region after the transition (figure \ref{fig:transtion_W=0.7}). We find that changes in Chern number are accompanied by changes in the band structure which mirror that sketched in figure \ref{fig:butterfly_trans_schem}. We also find that, throughout the spectrum, band gaps with this structure have Chern numbers $\sigma_{1}$ and $\sigma_{2}$ which are separated by multiples of $q$; $\sigma_{2} - \sigma_{1} \in q \mathbb{Z}$.

In figure \ref{fig:transition_W} $q = 1$ and the statement that $\sigma_{2} - \sigma_{1} \in q \mathbb{Z} = \mathbb{Z}$ is trivial. There are non-trivial examples however. The structure sketched in figure \ref{fig:butterfly_trans_schem} can be seen elsewhere in the spectrum (Fig. \ref{fig:butterfly}) at non-unit values of $q$. For example: at $\phi / \phi_{0} = 2.\dot{3} = 7 / 3, \ 2.\dot{6} = 8 / 3, \ 3.\dot{3} = 10 / 3$, and $3.\dot{6} = 11 / 3$ we have $\sigma_{1} = -1$ (green) and $\sigma_{2} = 2$ (purple). In each case $q = 3$ and $\sigma_{2} - \sigma_{1} = 3$. As another example take $\phi / \phi_{0} = 2.75 = 11 / 4$ and $3.75 = 15 / 4$. Here $q = 4$ and $\sigma_{1} = -1$ (green) and $\sigma_{2} = 3$ (pink) and $\sigma_{2} - \sigma_{1} = 4 = q$. We have also verified this for $\phi / \phi_{0} = 2.4 = 12 / 5$ and $2.6 = 13 / 5$ (difficult to see in Fig. \ref{fig:butterfly}) where $\sigma_{1} = -2$ (yellow) and $\sigma_{2} = 3$ (pink) so that $\sigma_{2} - \sigma_{1} = 5 = q$.

The Diophantine equation (\ref{equ:diophantine}) of Ref.\cite{thoulessQuantizedHallConductance1982} provides some context for this observation. As was seen above, for a given energy gap the possible values of $\nu$ are separated by multiples of $q$. Therefore, if the Chern number was to change as in figure \ref{fig:butterfly_trans_schem} then the original value, $\sigma_{2}$, must be separated from the final value, $\sigma_{1}$, by a multiple of $q$. Seeing only the initial or the final configuration depicted in figure \ref{fig:butterfly_trans_schem} (i.e. with $\sigma_{2} - \sigma_{1} = nq$) suggests that a transition is possible.

\section{Spatial Distribution of the Current Density}

In this section we present our results for the spatial distribution of the current density over the sample. The central message is that introduction of the AG potential leads to a significant bulk current. We consider how the current depends on the position of $\mu$ within the Hofstadter butterfly. The values of $\mu$ considered are those labelled in figure \ref{fig:disp_phi=3.0} and the parameters used to compute current density are identical to those in figure \ref{fig:disp_phi=3.0} except for $L_{x}$. Since the calculation of current density is computationally intensive we reduce the system size to $L_{x} = 10a$. The effect of this is only to scale the bulk density of states and to change the shape of the edge bands (whilst keeping the Chern number unchanged).

\begin{widetext}
\begin{figure}
\centering
\subfloat
    [\label{fig:jyMapMu1}]
    {\includegraphics[width=0.5\textwidth,page=1]{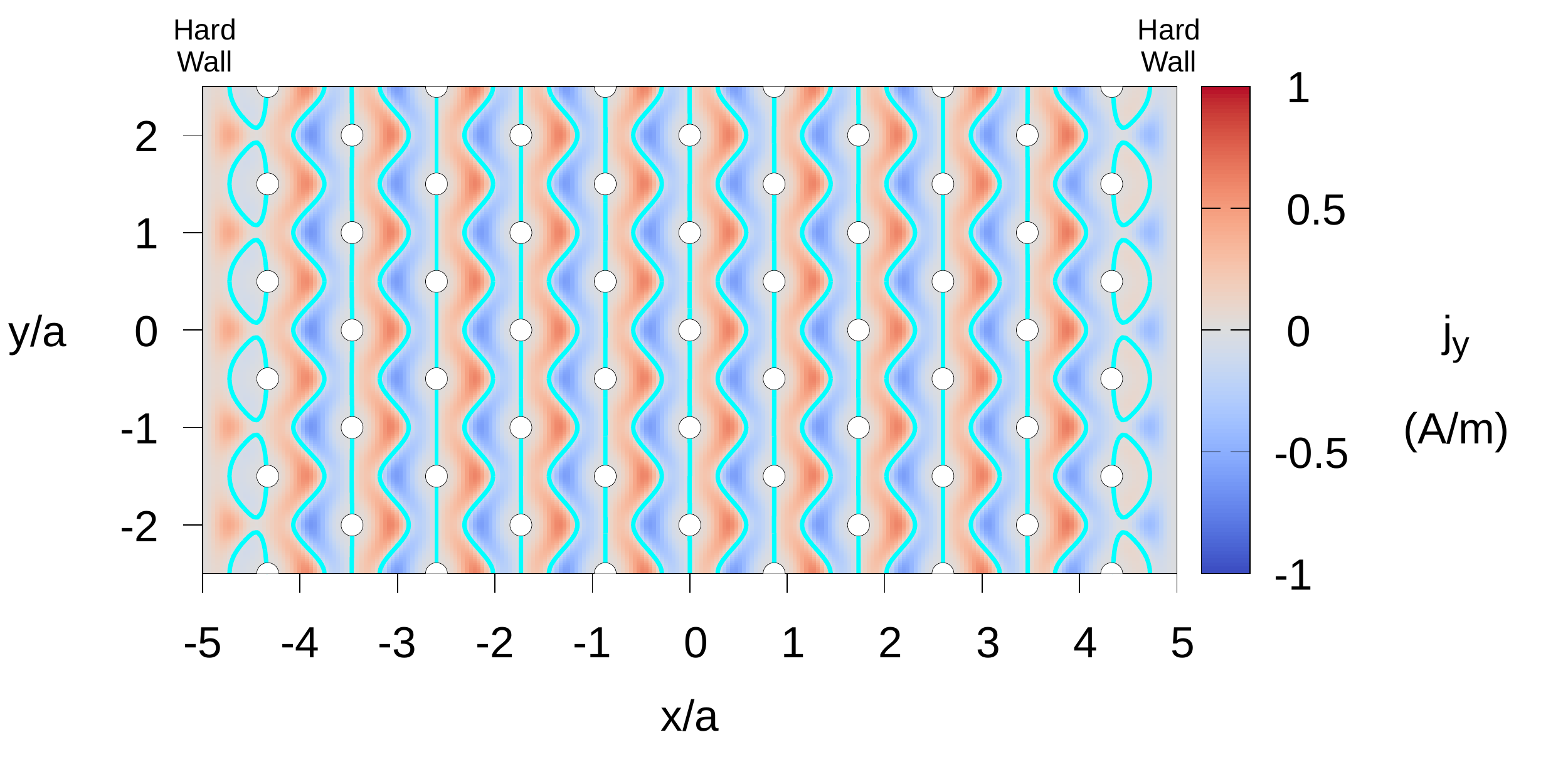}}
\subfloat
    [\label{fig:jyCutMu1}]
    {\includegraphics[width=0.4\textwidth]{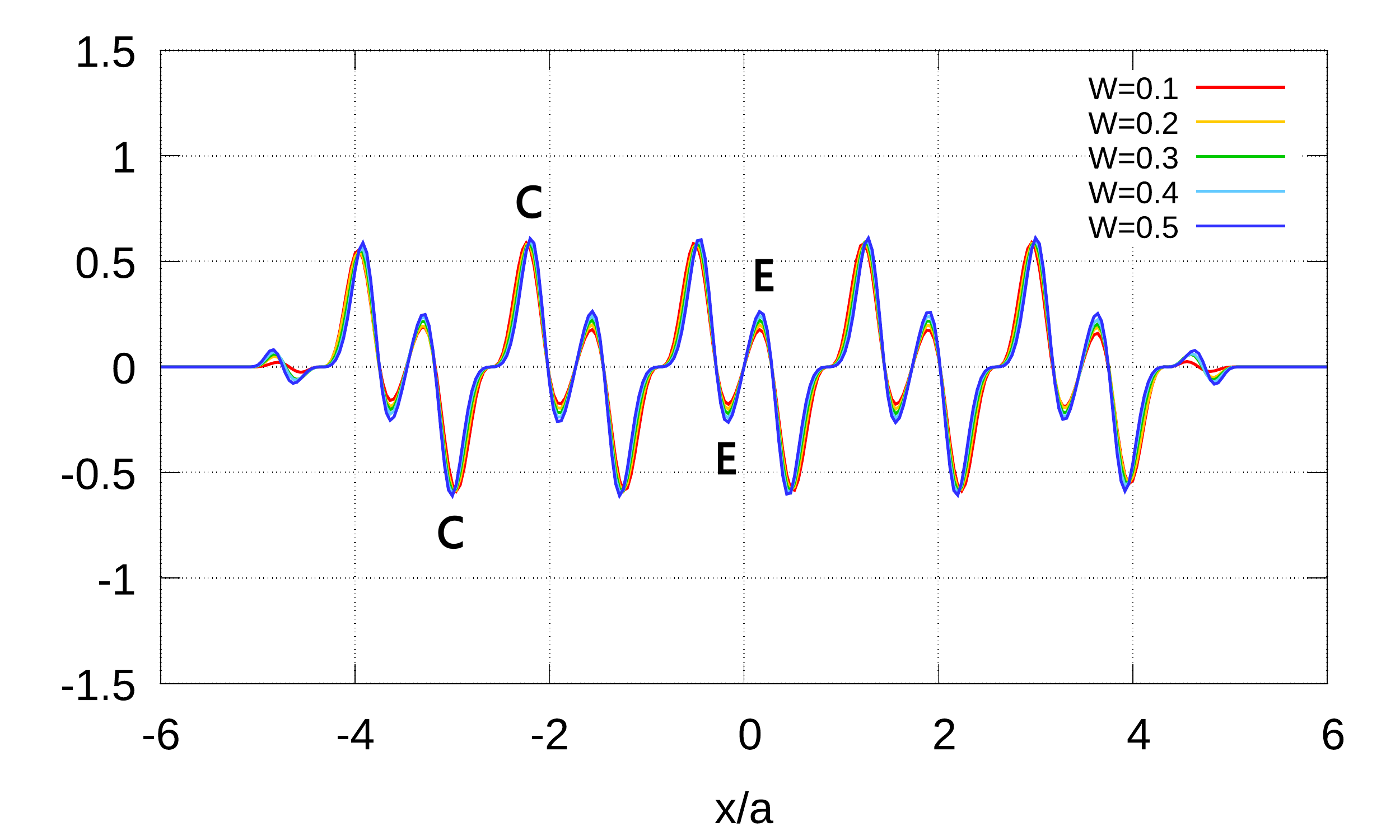}}

\subfloat
    [\label{fig:jyMapMu2}]
    {\includegraphics[width=0.5\textwidth,page=2]{{phi=3.0_W=0.4_map_jy}.pdf}}
\subfloat
    [\label{fig:jyCutMu2}]
    {\includegraphics[width=0.4\textwidth]{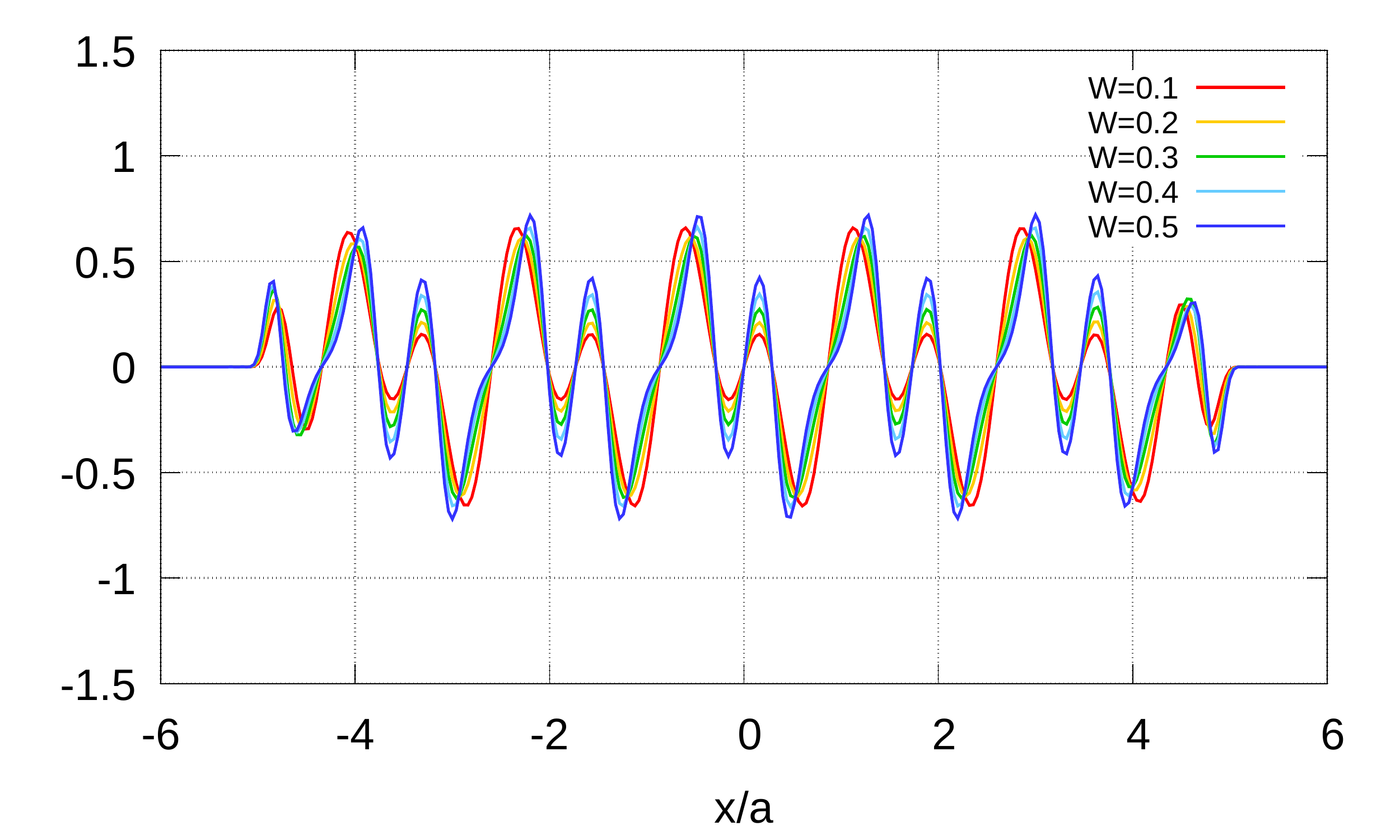}}

\subfloat
    [\label{fig:jyMapMu3}]
    {\includegraphics[width=0.5\textwidth,page=3]{{phi=3.0_W=0.4_map_jy}.pdf}}
\subfloat
    [\label{fig:jyCutMu3}]
    {\includegraphics[width=0.4\textwidth]{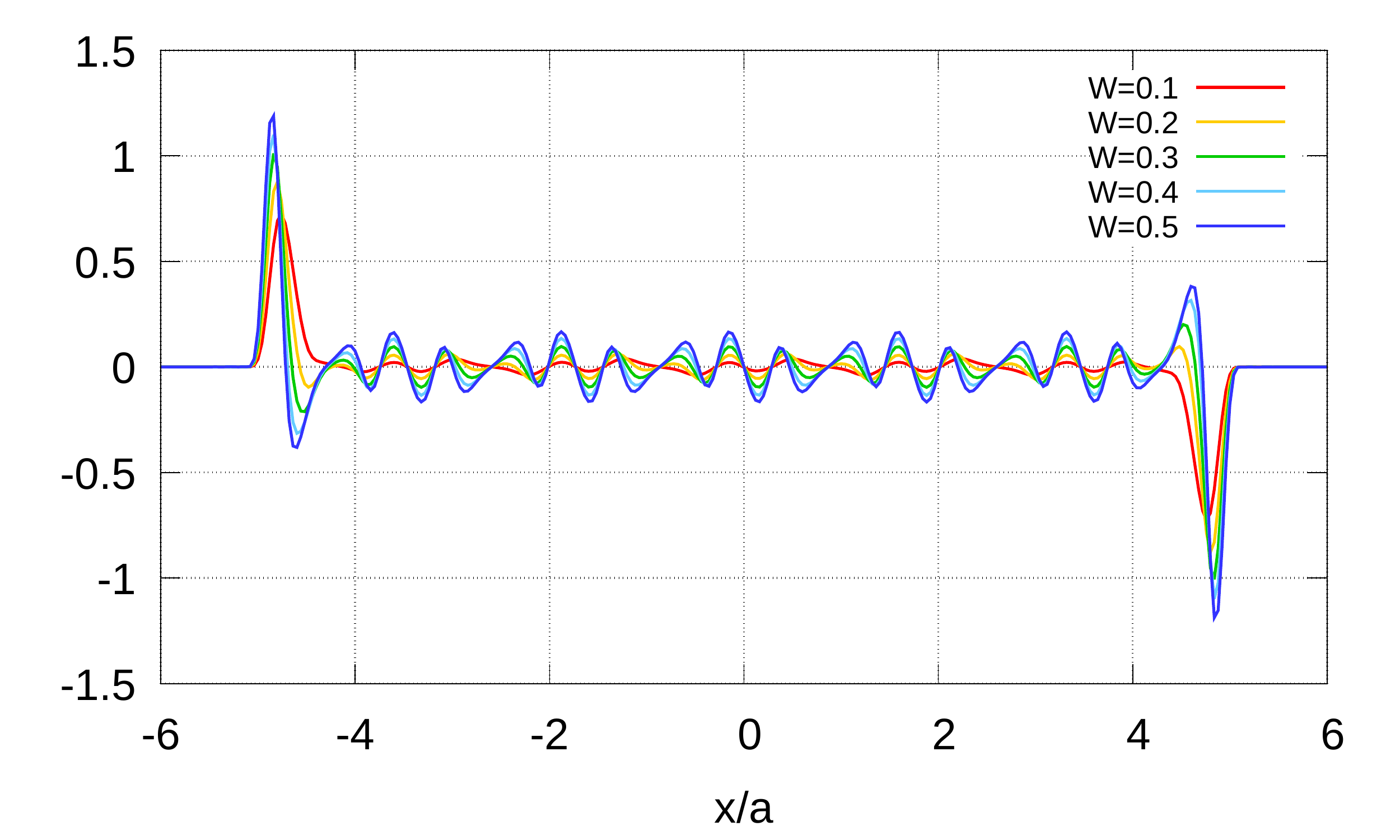}}
\caption{Overview of computed $j_{y}$ data at each value of the chemical potential shown in figure \ref{fig:disp_phi=3.0}. The first column ((a), (c), (e)) contains colour maps of $j_{y}(x,y)$ at $W = 0.4$ meV, $a = 120$ nm and $\phi / \phi_{0} = 3$ for (a) $\mu = \mu_{1}$, (b) $\mu = \mu_{2}$ and (c) $\mu = \mu_{3}$ (see figure \ref{fig:disp_phi=3.0}). Solid blue lines indicate the points for which $j_{y}(x,y) = 0$ and white circles indicate anti-dot lattice sites. Each plot in the second column ((b), (d), (e)) shows a cut of the data to its left taken at $y = 0.5a$ (i.e. $j_{y}(x,0.5a)$). In addition we include the data for $W = 0.1, \ 0.2, \ 0.3, \ 0.5$ meV. The labels $C$ and $E$ in (b) indicate the circulating and extended parts of the current respectively. In each calculation we fix $L_{x} = 10 a$.}
\label{fig:jy_all}
\end{figure}
\end{widetext}

Figure \ref{fig:jy_all} contains our results for the current density. The first column in figure \ref{fig:jy_all} presents colour maps of the $y$ component of the current density at $\mu_{1}$, $\mu_{2}$ and $\mu_{3}$ for $W = 0.4$ meV and $\phi = 3\phi_{0}$ (as in figure \ref{fig:disp_phi=3.0}). Zero current contours are drawn in light blue. We have labelled each lattice site with a white circle and have placed the boundaries of the plot at the hard walls of the confining potential (\ref{equ:edge_potential_function}). Each plot in the second column of figure \ref{fig:jy_all} shows a horizontal cut of the colour map to its left taken along $y = 0.5a$.

Our first observation is that current in the bulk of the sample is arranged into a set of extended streams which pass through the entire sample. This follows from the fact that zero current contours form open regions which connect the top and bottom of the plotted area and that the current is periodic along the $y$-direction. Each stream that carries current along a particular direction is be balanced by a stream carrying current along the opposite direction. We observe that these streams are present in every energy gap and at all magnetic field values tested. As will be discussed below, the magnitude of the extended streams of current is greatly reduced when moving from a fractionally filled Landau level to a fully filled Landau level.

On top of the extended currents there is a second component which circulates around anti-dots; in figures \ref{fig:jyMapMu1} and \ref{fig:jyMapMu2} this current is counter-clock-wise. We have verified this interpretation by looking at the $x$ component of the current. These two different components are labelled in figure \ref{fig:jyCutMu1}. An example of extended streams in this plot are the two small peaks centered around $x = 0$ and labelled `$E$'. And an example of circulating currents are the two larger peaks centered at $x = - 3 \sqrt{3} a / 2 \approx -2.6 a$ labelled `$C$'. This value of $x$ corresponds to a lattice site.

The current profile evolves as the chemical potential is moved through a Landau level. A current distribution is established after the chemical potential moves through the first subband in figure \ref{fig:disp_phi=3.0} to $\mu_{1}$ (figure \ref{fig:jyCutMu1}). When $\mu$ increases to $\mu_{2}$ the amplitude of this distribution increases slightly (figure \ref{fig:jyCutMu2}). A further increase of $\mu$ to $\mu_{3}$, so that all states within the Landau level contribute to the current density, kills most of the bulk current (figure \ref{fig:jyCutMu3}). This evolution between $\mu_{2}$ and $\mu_{3}$ is general, it applies at all the magnetic field values tested. Moving the chemical potential across the final subband always reduces the bulk current to a fraction of its original value. As will be discussed below this fraction is dependent on $W$.

\subsection{Scaling of Currents with Potential Strength}

The bulk distribution of current is dependent on potential strength $W$. In figures \ref{fig:jyCutMu1}, \ref{fig:jyCutMu2} and \ref{fig:jyCutMu3} we consider the current density at $W = 0.1$, 0.2, 0.3, 0.4 and 0.5 meV.

The bulk current scales with $W$ in different ways depending on the value of $\mu$. At $\mu = \mu_{3}$ (figure \ref{fig:jyCutMu3}) the bulk current scales roughly in proportion to $W$. This is consistent with the bulk current being zero in the $W \rightarrow 0$ limit. This proportionality was observed at all magnetic fields tested when $\mu$ covers the entire Landau level. For values of $\mu$ \textit{inside} the Landau level the scaling is more complicated. In figure \ref{fig:jyCutMu1} ($\mu = \mu_{1}$) both the extended and circulating parts of the bulk current are independent of $W$. In the limit $W \rightarrow 0$ the bandwidth of the Landau level goes to zero and $\mu_{1}, \ \mu_{2}, \ \mu_{3}$ approach the same value. It then becomes unclear how to define a fractional filling of the Landau level. For small (but finite) $W$, however, we observe little or no scaling in the bulk current at $\mu_{1}$. When $\mu$ is increased to $\mu_{2}$ (figure \ref{fig:jyCutMu2}) the circulating part of the current density remains independent of $W$ while the extended streams scale weakly with $W$. At other magnetic field strengths the scaling of bulk current density with $W$ follows a similar pattern.

Our results seem to suggest a non-zero bulk current exists for a weak lattice potential. This does not contradict the fact that for no lattice at all the current due to an entire Landau level is zero everywhere in the bulk. Bulk current is zero for a fully filled Landau level because the sum of all Larmor orbits (enumerated by their center coordinate $x_{k}$) cancel each other out. A small perturbation (i.e. a lattice potential) gives some small broadening to each Landau level. It is then possible to have fractional filling factors since the huge Landau degeneracy is lifted. Fractional filling allows for a current distribution due to some subset of all Larmor orbits which do not fully cancel each other and therefore produce a non-zero bulk current. Our statement is that this nonzero bulk current is fully pinned to the anti-dot sites.

\section{Conclusions}

We have studied the Chern numbers and equilibrium bulk/edge current density of electrons in a two-dimensional triangular lattice with out of plane magnetic field. These results derive from the eigenvalues and eigenstates of the corresponding single particle Hamiltonian, which we compute numerically.

In experimental realisations of AG with tunable potential strength it is useful to calibrate the modulation strength and determine which values of $W$ are being accessed. Observation of the Hall plateau transitions discussed in this work could provide a way to pin down the experimental value of $W$. Such an experiment would be limited by the small energy gap sizes ($\lesssim 0.5$ meV).

In principle, the equilibrium current distribution is measurable through the magnetic field it produces by NV centre magnetometry \cite{taylorHighsensitivityDiamondMagnetometer2008} \cite{hongNanoscaleMagnetometryNV2013}. We estimate magnetic fields on the order of 0.1 milligauss at 100 nm above the 2DEG which is well within the sensitivity of this method. The limiting factors are two fold. Firstly the spatial resolution which would need to be $\lesssim$ 10 nm. And secondly one would need to detect a $0.1$ miligauss signal in the presence of a background field on the order of 1 T.

\section*{Acknowledgements}

We acknowledge useful discussions with J. Cole, A. R. Hamilton, O. Klochan, Y. Lyanda-Geller, A. H. MacDonald, O. A. Tkachenko, V. A. Tkachenko, G. S. Uhrig and D. Wang. This research was supported by an Australian Government Research Training Program (RTP) Scholarship and includes computations using the computational cluster Katana supported by Research Technology Services at UNSW, Sydney. We have also received support from the Australian Research Council Centre of Excellence in Future Low-Energy Electronics Technology (CE170100039).

\bibliography{bibliography.bib}

\end{document}